\documentclass{aastex701}

\begin{document}

\title{Observations with the Southern Connecticut Stellar Interferometer. 
II. First Three-Telescope Observations and a New Diameter Measurement of Arcturus}

\author[orcid=0000-0003-2159-1463,sname='Horch']{Elliott P. Horch}
\altaffiliation{Adjunct Astronomer, Lowell Observatory.}
\affiliation{Department of Physics, Southern Connecticut State University, 501 Crescent Street,
New Haven, CT 06515, USA}
\email[show]{horche2@southernct.edu}  

\author[sname='Lucero']{Sebastian M. Lucero} 
\affiliation{Department of Physics, Southern Connecticut State University, 501 Crescent Street,
New Haven, CT 06515, USA}
\email{luceros2@southernct.edu}

\author[orcid=0000-0001-8692-501X,sname='Martone']{Max Martone}
\affiliation{Department of Physics, Southern Connecticut State University, 501 Crescent Street, 
New Haven, CT 06515, USA}
\email{martonem5@southernct.edu}

\author{Riley C. Barrett}
\affiliation{Department of Physics, Southern Connecticut State University, 501 Crescent Street, 
New Haven, CT 06515, USA}
\email{barrettr6@southernct.edu}

\author[orcid=0009-0005-2867-3180,sname='Baculima']{Ana I. Baculima Dur\'{a}n}
\affiliation{Department of Physics, Southern Connecticut State University, 501 Crescent Street, 
New Haven, CT 06515, USA}
\email{baculimaa1@southernct.edu}

\author[orcid=0000-0002-1155-977X,sname='Powers Ozyurt']{Fiona T. Powers \"{O}zyurt}
\affiliation{Department of Physics, Southern Connecticut State University, 501 Crescent Street, 
New Haven, CT 06515, USA}
\email{powersozyuf1@southernct.edu}

\author[orcid=0009-0005-7107-4968,sname='Posick']{Gage Posick}
\affiliation{Department of Physics, Southern Connecticut State University, 501 Crescent Street, 
New Haven, CT 06515, USA}
\email{posickg1@southernct.edu}

\author[orcid=0009-0007-7231-5529,sname='Petroski]{Alexander Petroski}
\affiliation{Department of Physics, Southern Connecticut State University, 501 Crescent Street, 
New Haven, CT 06515, USA}
\email{petroskia1@southernct.edu}

\author[orcid=0009-0007-1284-7240]{James W. Davidson, Jr.}
\affiliation{Department of Astronomy, University of Virginia, 530 McCormick Road, Charlottesville, VA 22904, USA}
\email{jimmy@virginia.edu}

\author[orcid=0000-0003-2025-3147,sname='Majewski]{Steven R. Majewski}
\affiliation{Department of Astronomy, University of Virginia, 530 McCormick Road, Charlottesville, VA 22904, USA}
\email{srm4n@virginia.edu}

\author[orcid=0009-0005-6572-3853,sname='Pellegrino']{Richard A. Pellegrino}
\affiliation{CSCU Center for Quantum and Nanotechnology, Southern Connecticut State University, 501 Crescent Street, 
New Haven, CT 06515, USA}
\email{pellegrinoa2@southernct.edu}

\author{Paul M. Klaucke}
\altaffiliation{Current address: Department of Physical and Life Sciences, Quinnipiac University, 275 Mount Carmel Avenue,
Hamden, CT 06518 }
\affiliation{Department of Physics, Southern Connecticut State University, 501 Crescent Street, 
New Haven, CT 06515, USA}
\email{Paul.Klaucke@quinnipiac.edu}

\author[orcid=0009-0000-5136-6924,sname='Lesley']{Xavier Lesley-Salda\~{n}a}
\altaffiliation{Current address: Department of Astronomy, The Ohio State University, 140 West 18th Avenue,
Columbus, OH 43210, USA}
\affiliation{Department of Physics, Southern Connecticut State University, 501 Crescent Street, 
New Haven, CT 06515, USA}
\email{lesley.10@buckeyemail.osu.edu}

\author[orcid=0009-0007-4277-0360,sname='Sutherland]{Torrie Sutherland}
\altaffiliation{Current address: Apache Point Observatory, P.O. Box 59, Sunspot, NM 88349-0059}
\affiliation{Department of Physics, Southern Connecticut State University, 501 Crescent Street, 
New Haven, CT 06515, USA}
\email{torrieps@nmsu.edu}

\author[orcid=0000-0002-5870-8488,sname='Weiss']{Olivia S. Weiss}
\altaffiliation{Current address: Lowell Observatory, 1400 West Mars Hill Road, Flagstaff, AZ 86001}
\affiliation{Department of Physics, Southern Connecticut State University, 501 Crescent Street, 
New Haven, CT 06515, USA}
\email{sweiss1993@gmail.com}

\begin{abstract}

We discuss the most recent observations made with the Southern Connecticut Stellar Interferometer
(SCSI), which is a three-station stellar intensity interferometer located on the campus of Southern Connecticut
State University, in New Haven, Connecticut. Two different kinds of observations are presented. We first analyze 
observations of Vega taken in a three-telescope mode. (Previously, the instrument had only two operational stations.)
We show that, while the efficiency remains nearly identical to that reported in our last paper, the addition of the third station
allows more photon data to be recorded simultaneously, and therefore we can build up the photon-bunching peak 
in the data stream in fewer hours on sky for an unresolved source. In the second part of the paper, we report our observations to date of
the nearby red giant star, Arcturus, most of which occurred in the first half of 2025. 
These show that, as a partially resolved source at the baselines we used, we detect fewer
correlations in the photon-bunching peak than for an unresolved source of comparable brightness. 
Combining the data with speckle imaging observations taken at Apache Point Observatory, we derive a new measurement 
of Arcturus' diameter that extends the time baseline of interferometric observations of the star and 
is consistent with previous analyses made by other investigators. 

\end{abstract}

\keywords{\uat{Astronomical instrumentation}{799} --- \uat{Interferometers}{805} --- \uat{Field stars}{2103} --- \uat{Stellar radii}{1626}}


\section{Introduction} 

Stellar intensity interferometry (SII), which was first discussed by  \citet{1956Natur.177...27B, 1958RSPSA.248..199B}, 
is a technique used for deriving 
extremely high-resolution information of astronomical sources without the need of interfering the light from different stations. Instead, correlated intensity variations are observed when the frequency bandwidth of the light detected is 
narrow enough to reveal non-Poissonian fluctuations in the photon flux received at each telescope, known as the Hanbury Brown and 
Twiss (HBT) effect, or, in the photon counting regime, photon bunching. The degree of correlation observed at a given baseline can
then be used to infer a Fourier component of the spatial distribution of the source, in other words, to make a visibility measurement. For visible light, the degree of correlation is usually very small, and so what limits the technique is the 
signal-to-noise ratio that is achievable; nonetheless, Hanbury Brown and his 
collaborators succeeded in building a two-station interferometer with very large telescopes. Details of the astronomical observations done with this instrument
appear in \citet{1974MNRAS.167..121H}. Even with the large size of the collecting apertures, the instrument was only capable
of making diameter measurements of very bright stars, ostensibly due to the timing precision of the photmultiplier tubes
and electronics available at the time. Because of this, SII fell out of favor as the technical feasibility of 
high-precision optical metrology was realized and the research
behind the current generation of long baseline optical interferometers in astronomy gained momentum in the 1980's.

However, there is a renewed and growing interest in SII today, and several groups 
have successfully revived and improved the technique with modern instrumentation. 
Generally, the systems used in these projects have fallen into two very different categories. In the first category, large telescopes that 
do not have optical-grade image quality are used, and the light is focused into a relatively large spot ($\sim$1cm in diameter), where it 
strikes a photomultiplier tube (PMT) and is read out there. These are in effect a repurposing of 
Imaging Atmospheric Cherenkov Telescopes (IACTs) for periods when the Moon is too bright to observe Cherenkov
radiation from energetic particles entering the atmosphere.
This approach is similar to the approach of Hanbury Brown and Twiss, but with
the advantage of modern photmultiplier tubes, which have substantially higher quantum efficiencies and better timing characteristics 
than those of the 1970's.
The Very Energetic Radiation Imaging Telescope Array System (VERITAS) \citep{2020NatAs...4.1164A}, 
the Major Atmospheric Gamma-Ray Imaging Cherenkov (MAGIC) array \citep{2020MNRAS.491.1540A}, 
and High Energy Stereoscopic System (H.E.S.S.) \citep{2024MNRAS.52712243Z} are examples of this 
type of system. The other approach, which includes the Arizona State University Stellar Intensity Interferometer (ASUSII) \citep{2025MNRAS.537.2527M}, the Nice Observatory system \citep{2018ExA....46..531R}, 
the Asiago Observatory system \citep{2021MNRAS.506.1585Z},
and our own work  \citep{2022AJ....163...92H}, hereafter Paper I, is to use smaller 
telescopes that deliver true optical quality, and to focus the relatively small spot of light onto a very fast detector, typically a 
single-photon avalanche diode (SPAD) detector with timing jitter of less than 100 ps.

Most recently, some of these groups have gone beyond the detection of photon bunching to making increasingly 
high-precision measurements of stars.
\citet{2020MNRAS.494..218R} presented an analysis of P Cyg where SII data was combined with high-resolution 
spectroscopy to infer the distance to the star; the result was $1.56 \pm 0.25$ kpc, which is comparable in precision to {\it Gaia}.
Both \citet{2022MNRAS.515....1D} and \citet{2023AJ....165..117M} followed that work with studies of $\gamma$ Cas and
Rigel.
\citet{2024ApJ...966...28A} measured the diameter of $\beta$ UMa at visible wavelengths for
the first time, obtaining a precision of roughly 5\%, 
\citet{2025arXiv250615027A} has recently measured the oblateness of $\gamma$ Cas, 
also to a few percent, and 
 \citet{2025MNRAS.537.2334V} detail new measurements of the diameters of four 
stars ($\beta$ Cru, $\eta$ Cen, $\sigma$ Sco, and $\delta$ Sco) with uncertainties ranging from 2 to 8\%. 
These
results, while not yet as precise as what is currently possible with Michelson-style long baseline optical interferometers, 
nonetheless show
the potential of the modern SII today. This has in turn spurred on investigations to improve 
efficiency in such systems, examples of which include \citet{2024sf2a.conf..197T}, \citet{2025JATIS..11c5005L}, \citet{2025SPIE13448E..02K}, 
\citet{2025MNRAS.538..867M}, and \citet{2025arXiv251116505C}.

In the present study, we build upon the work in our Paper I to do two things. 
First, we detect the HBT signal with SCSI using three telescopes for the first time, 
allowing us to refine our understanding of the instrument performance. We then 
describe SCSI observations of Arcturus ($= \alpha$ Boo, HR 5340, HIP 69673), which show partial correlation at baselines of 2-3 m. 
Combining with speckle observations taken with the Differential Speckle Survey Instrument (DSSI) at Apache Point Observatory, 
we are able make a new diameter measurement of the star. We discuss this result in light of previous work.

\section{Three-Telescope Operations}

\subsection{Instrumentation and Observing Method}

In Paper I, we describe our instrumentation and the observing process for a two-telescope
arrangement. The main components used in this mode are (1) two 61-cm portable Dobsonian 
telescopes made by Equatorial Platforms, (2) MicroPhotonDevices PDM Series SPAD detectors, and (3) a PicoQuant PicoHarp 300 
timing module. The SPAD detectors are connected to an optical harness that contains a collimating lens, the
narrow bandpass filter (usually 532 nm with a full width at half maximum [FWHM] of 1.2 nm) and
a reimaging lens that focuses the light onto the detector. This detector package is
then mounted at the Newtonian focus of each telescope, and 
signal wires are connected from the SPADs to the two input channels of the timing module. The 
PicoHarp has the ability to timestamp the detected photons with a precision of 4 ps; the detector timing
jitter, however, results in cross-correlation timing precision of $\sim$50 ps, and so, as described in Paper I, 
we typically bin the data into
64-ps bins in order to identify correlation peaks. In this case, the HBT peak is effectively a small delta
function sitting on top of the constant pedestal of random correlations in the timing cross-correlation function
made from the photon timings recorded at each station.

In the three-telescope observations presented here, a third identical telescope and detector are added to the 
set-up, and we use an 8-channel correlator, the PicoQuant HydraHarp 400 (where five of the eight available channels
are not used). The HydraHarp can time stamp events to 1 ps, and again, this is much higher than the 
detector timing jitter. Telescopes 1 and 2 were set up in exactly the
same way as in Paper I; they were aligned in a north-south direction with Telescope 2 north of 
Telescope 1, based on paint spots
on the asphalt at the observing location that were originally marked by observing the shadow cast by a plumb bob 
at solar noon several years ago and repainted
twice since then. Telescope 3 was then set up west of Telescope 1, so that the locations of the telescopes were at
the vertices of a right triangle (with Telescope 1 at the apex of the 90-degree angle). The legs of the right triangle 
were then nominally along the north-south and east-west directions, and the hypotenuse was represented by the 
baseline between Telescopes 2 and 3. Generally, we tried to place Telescope 3 in a position where the east-west 
baseline was approximately the same length as the north-south baseline. The placement of Telescope 3 was aided 
by long wooden beams that could easily be placed on the asphalt in orthogonal directions. These were laid along 
both the north-south paint spots and in the orthogonal direction. For each observing session,
photographs were taken of the arrangement so that, if there were significant deviations from the north-south
and east-west directions, these could be estimated from these images. The length of the north-south and
east-west baselines was measured with a tape measure, using identical fiducial positions on each telescope.

Once the telescopes were placed and their positions documented, observing proceeded by first checking the collimation
of each telescope with a laser collimator and then performing pointing
alignments on two bright stars separated on the sky by at least 30 degrees, 
after which the telescopes would track. We then place the SPAD detectors at the focal
point, and whenever possible, we would first point toward a planet (usually Jupiter or Mars) to align the 
finder telescopes and the detectors. Because the detectors have a very small active area (a diameter of 100 microns) and
are not perfectly aligned with the optical axis, the technique of using an extended object allows us to quickly adjust the finder 
to an approximate match to the detector position in the field of view. Once roughly aligned, we would point to
the science target and attempt to maximize the count rate on each detector. When counts were maximized at all
telescopes, we took data in five-minute 
files, with an observer manually guiding each telescope throughout to keep the count rate as high as possible. 
(In five minutes, the shift of the cross-correlation peak within the data file was typically of the order of one 64-ps bin, 
and without manual guiding, 
we find that stars often slide off the detector in 10 to 30 seconds due to imperfect tracking.) Thus, in 3-telescope mode, 
an observing team of at least four people is needed,
one to guide at each telescope, and a fourth person to manage the data collection process.

In order to process the data correctly, two types of calibration measurements are important. These are (1)
estimates of height differences between the telescopes as a function of position, and (2) signal timing differences
due to slight differences in cable length and intrinsic differences in the timing that can be measured between the 
channels of the timing correlators we use. For height differences, we use a spirit level starting at the location of 
Telescope 1 and then build up the (differential) topology of the site by placing the level at different places in a raster
sequence over the area used for our observations. The results indicate that the asphalt has a crown near the middle of the 
surface and slopes down near the edges, as expected for what is essentially a service road leading to one of the science 
buildings on our campus. The procedure for determining intrinsic and cable timing difference is detailed in Paper I;
the basic method is to illuminate a single detector in the lab with a lamp and then 
split the signal cable so that two channels of the timing correlator receive the same photon stream.
By using different combinations of channels and cables, we can derive the delays for each cable and channel relative 
to each other. There could also be optical path length differences on the order of a few millimeters since each telescope 
is manufactured separately and they are therefore not exactly identical; however, we find that these 
are not large enough to change the location of the correlation peak when binning the data in 64-ps intervals. One centimeter
of optical path delay represents a timing difference of only 33 ps, which is typically about half the length of the timing
bin size we use for cross-correlations; however, we will return to this point in the analysis 
of the data presented later in the paper.

\subsection{Data Reduction Methodology}

For three-telescope work, much of our data reduction methodology remains the same as 
described in Paper I, where the data presented were solely from a two-telescope arrangement
that was oriented in a north-south baseline. Timing cross-correlations are computed for each 
5-minute sequence recorded. However, 
cross-correlation peaks shift position as an observation proceeds, and so to combine those properly, 
one must take account of where the peak in a
given cross-correlation is expected to appear as a function of time, based on sky position of the target and the baseline.
Even within a 5-minute file, small shifts occur. Our cross-correlation routine accounts for these and 
corrects all timing differences to the value expected at the start of the file, but
leaves the peak position otherwise unchanged. When sequentially recorded 
files are combined, the shift value at the start of the file is computed,
and that location is shifted to a nominal ``zero'' position in each case. 
Thus, when adding files, the peak builds up in a common location,
defined to be zero delay. In practice, this zero position is
in fact the middle of the zero bin for 64-ps binning in time; in other words, it is 32 ps from true coincidence,
so that, when the data are binned down to 64-ps intervals, all of the correlations can be expected to be in
the ``zero'' timing difference sample, which extends from 0 to 63 ps.

An important part of the analysis of the data presented
in Paper I was the correction made because of presumed slight differences in the baseline orientation
of the telescopes from night to night. This is done by taking the final combined cross-correlation for a given night
and recomputing it with baseline orientations sweeping through a small range near the nominal position angle value of
the baseline. If the peak is enhanced through this process, it is assumed that the actual placement of the telescopes 
was not exactly the north-south orientation we attempted to achieve but deviated from that slightly, 
usually no more than 0.1 to 0.2 degrees. In Paper I we show through simulation that this process results in peak values that more 
nearly reach the theoretical value, whereas if this is not done, SNR is sacrificed. 

The addition of a third telescope creates 
additional complexity in the data reduction process, but also presents an opportunity for additional constraints to be applied.
Our approach is as follows. We compute cross-correlations for each pair of telescopes, resulting in three different summed 
cross-correlations from the data for a given night. We then do the angle correction, but subject to the constraint that the 
three telescopes must form a triangle; that is, the three interior angles created by the baselines must sum to 180 degrees.
We enforce this constraint by allowing the north-south and east-west baselines to independently vary, but constrain the third baseline
(between Telescopes 2 and 3) based on the other two angles. For a source where a peak is expected, we combine all three
cross-correlations and vary over this two-dimensional parameter space to determine the final angles. The final pair of
angles is the closest local maximum to the nominal set of baselines. Figure 1 gives an example of the
cross-correlations before and after this process for one night of data. 

\begin{figure}[!t]
\figurenum{1}
\plottwo{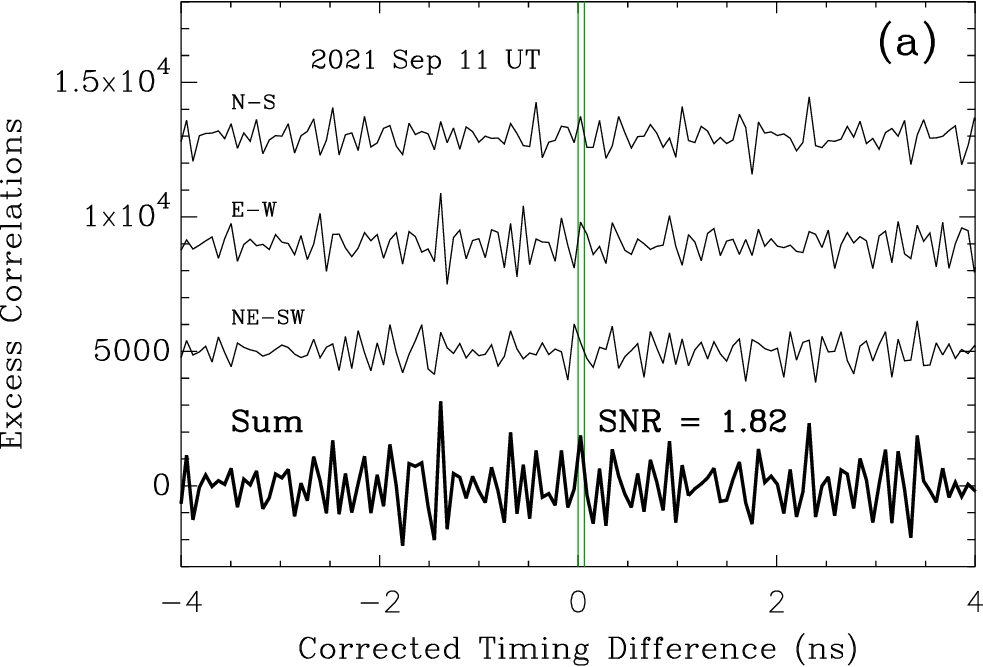}{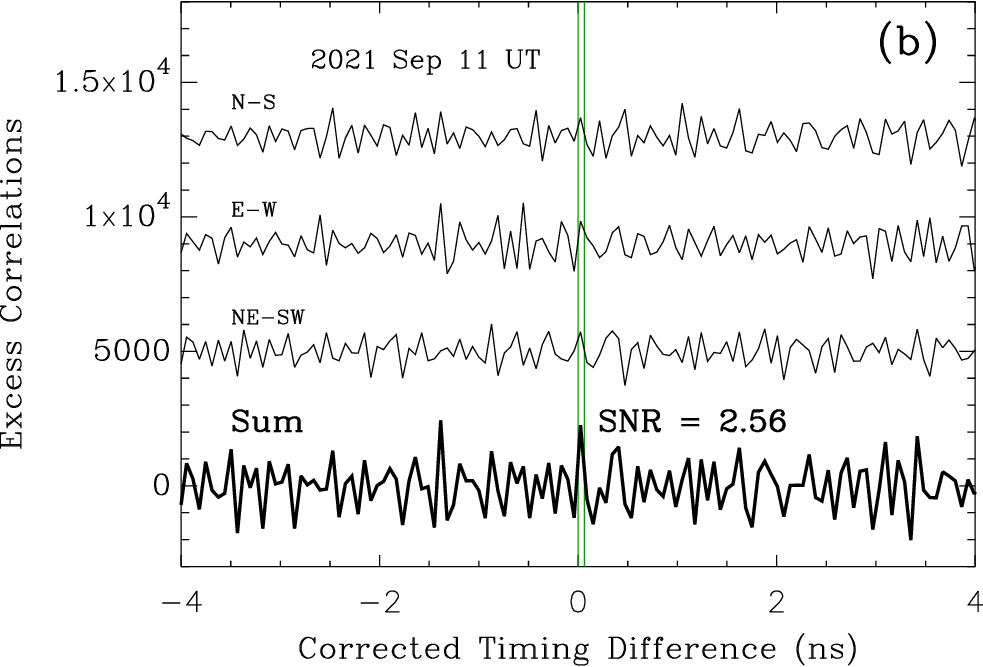}
\figcaption{The cross-correlation results for each baseline on 11 Sep 2021. (a) In this case, the plots are made
assuming the nominal position angles for the baselines (i.e. N-S = 180$^{\circ}$, E-W = 90$^{\circ}$, and NE-SW = the angle implied by 
completing the triangle with this baseline as the hypotenuse). (b) The same data, but using the method described in the
text for accounting for position angle offsets. In each plot, the data for the individual baselines are offset from the $y=0$ line
for clarity and the vertical green lines indicate the location of the expected cross-correlation peak.}
\end{figure}

\begin{deluxetable}{ccccccc}
\tablewidth{0pt}
\tablenum{1}
\tablecaption{New SCSI Observations of Vega.}
\tablehead{
\colhead{Date} & No. of &
\colhead{N-S Sep.} &
\colhead{E-W Sep.} &
\colhead{NE-SW Sep.} &
\colhead{Total Observing}&  \colhead{Avg. Count Rate}\\
(UT) & Telescopes\tablenotemark{a} & (m) &  (m) & (m) &Time (hr)\tablenotemark{b} & per Tel. (MHz)
}
\startdata
2021 Aug 25 & 3 & 2.529 & 2.651 & 3.664  & 0.42 & 0.150\\
2021 Sep 04 & 3 & 2.485 & 2.783 & 3.731 & 2.25 & 0.345\\
2021 Sep 11 & 3 & 2.454 & 2.452 & 3.469 & 3.34 & 0.709 \\
2022 Jun 29 & 3 & 2.528 & 2.388 & 3.478 & 1.00 & 0.211 \\
2025 May 12 & 2 & 2.151 & -- & -- &  1.17 & 0.878 \\
2025 May 27 & 2 & 2.224 & -- & -- & 0.92 & 1.311 \\
2025 Jun 04 & 2 & 2.061 & -- & -- & 2.08 & 0.625 \\
2025 Jun 25 & 3 & 3.403 & 2.955 & 4.507 & 1.41 & 0.098 \\
\enddata
\tablenotetext{a}{All observations taken with a filter with center wavelength of 532 nm and FWHM of 1.2 nm.}
\tablenotetext{b}{There are 8.42 hours of 3-telescope observations, 3.45 hours of 2-telescope 
observations, or 11.87 total hours of data obtained
on the eight nights represented here.}
\end{deluxetable}

\begin{figure}[!t]
\figurenum{2}
\plottwo{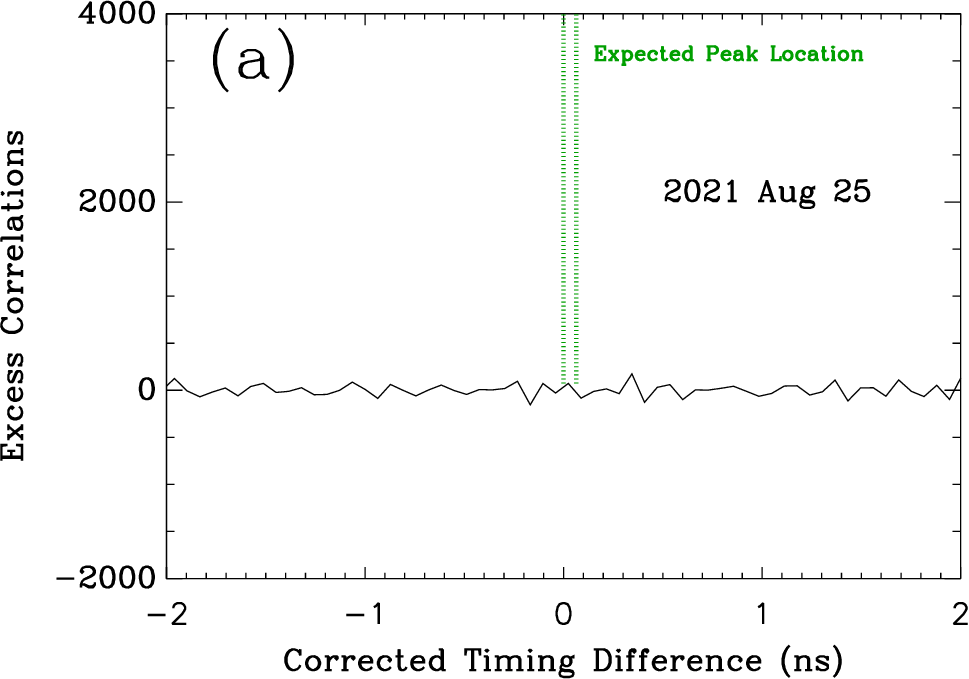}{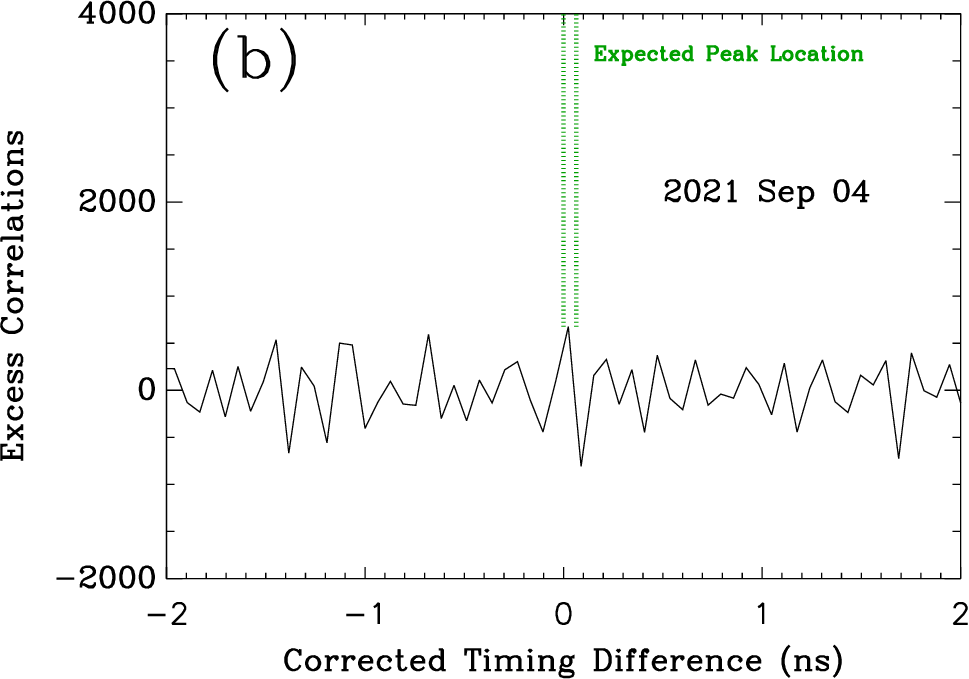}
\plottwo{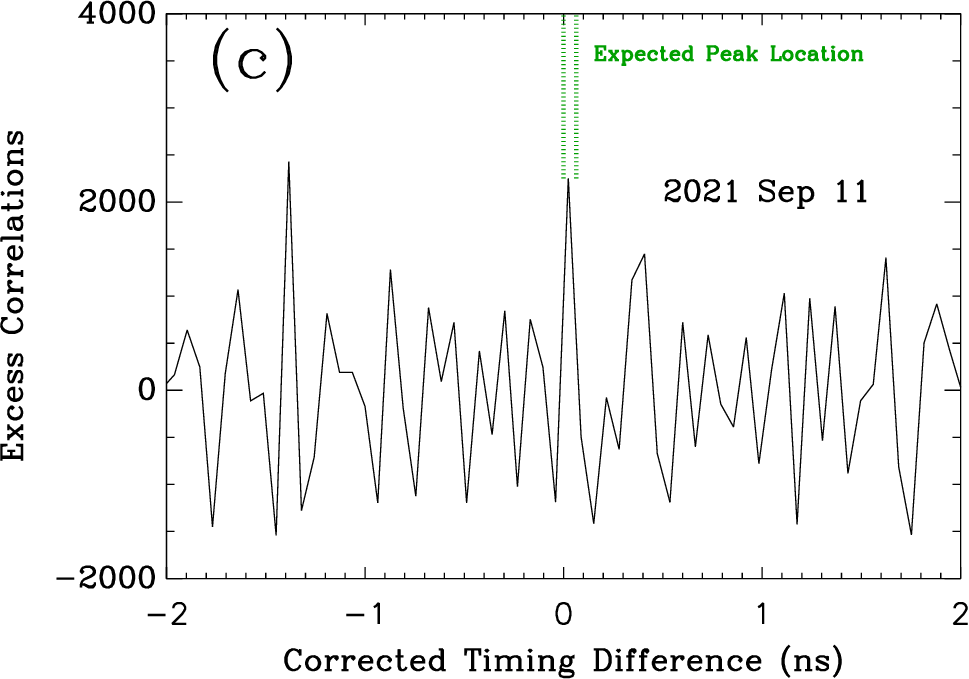}{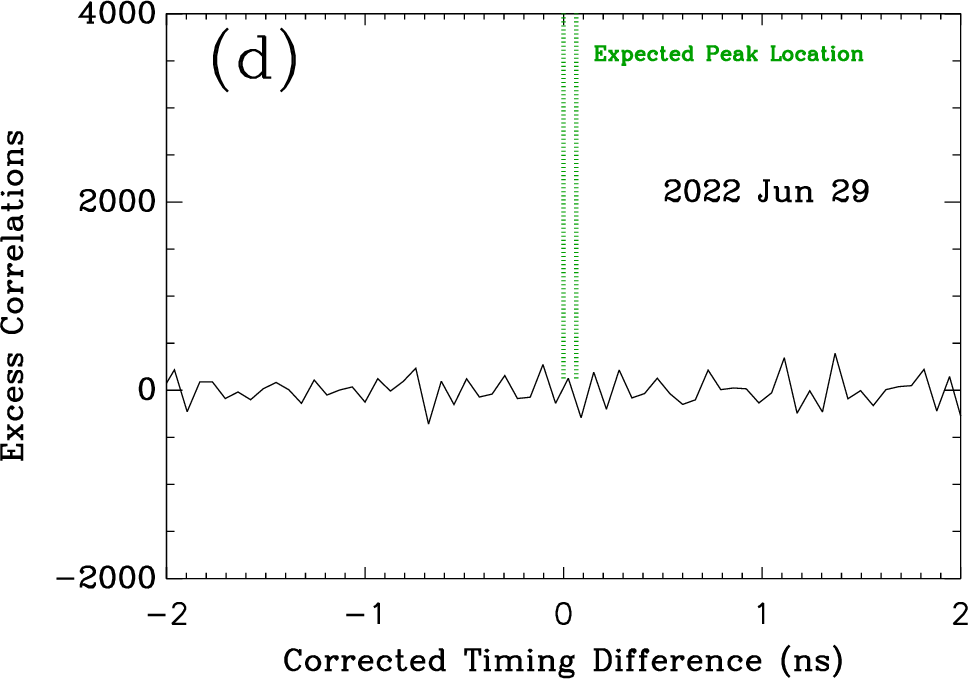}
\plottwo{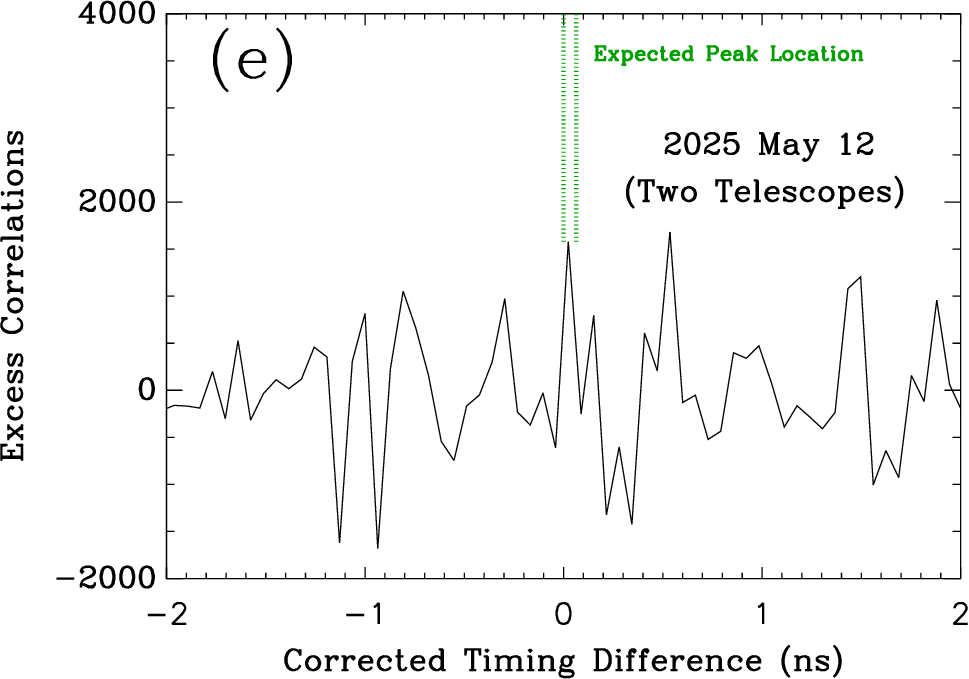}{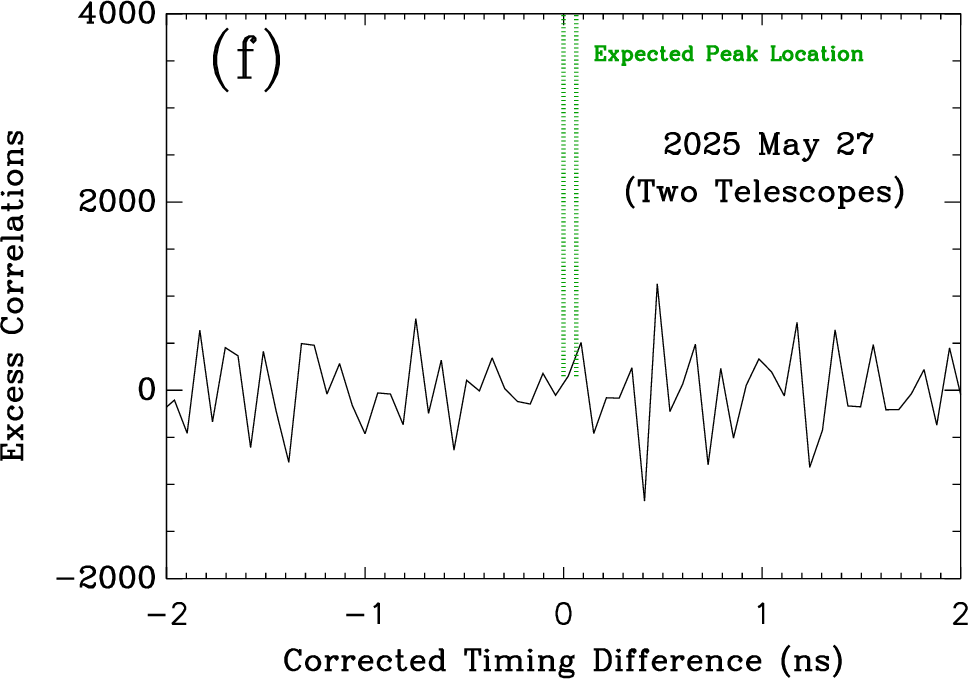}
\plottwo{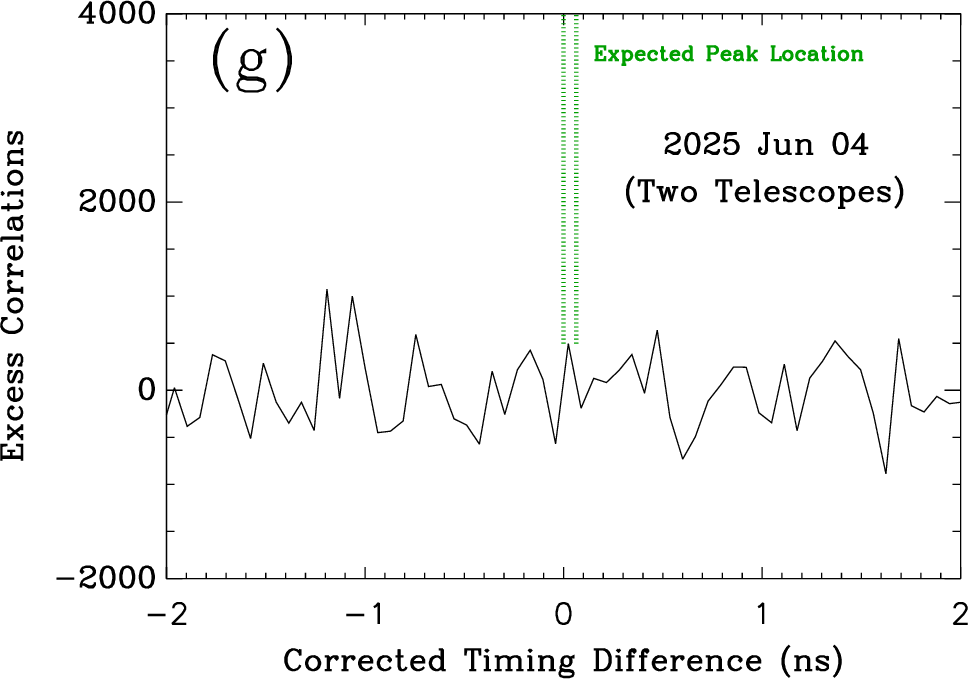}{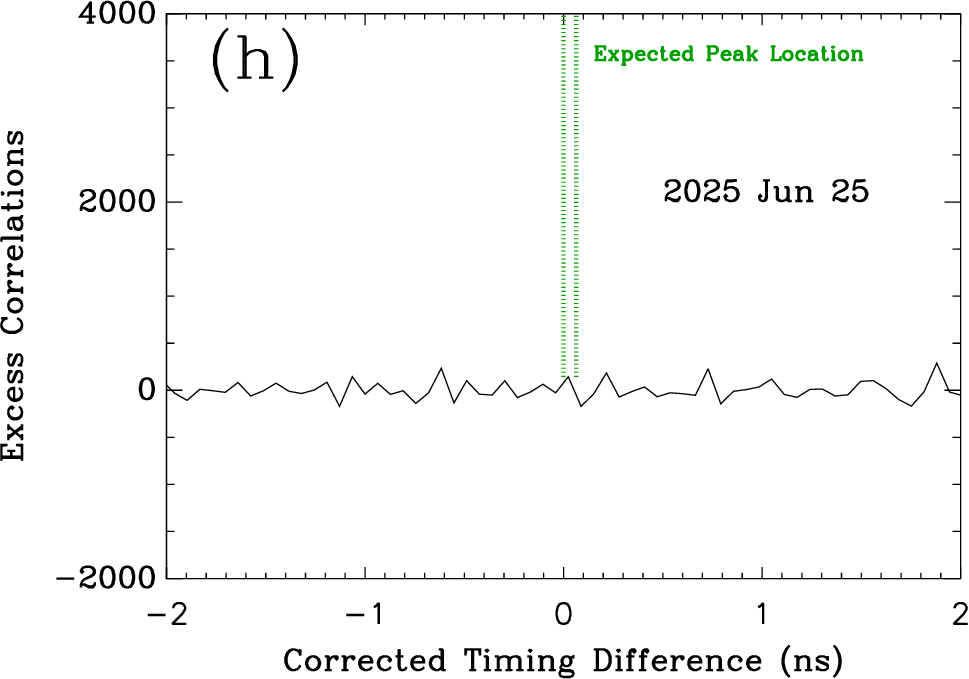}
\figcaption{Cross-correlations obtained for each night listed in Table 1. In each case, the mean number of correlations
per timing bin has been subtracted so that the HBT peak is more visible, and the green dotted lines mark the 
expected location of the peak. Except where otherwise noted, the three-telescope arrangement discussed in the text
was used.}
\end{figure}

\subsection{Three-Telescope Observations and Results}

We present in this section the results of five observing sessions of Vega in three-telescope mode, together with three 
nights of observations of the star in two-telescope mode taken on nights where we also observed Arcturus, as discussed 
in the next section. Table 1 details these observations, which in all cases were taken at baselines that were small enough that 
Vega would be unresolved, and the cross-correlation peak representing the HBT effect would be effectively 
maximized. This allows us to build up that peak as efficiently as possible, and to make a comparison between our results
here with those presented in Paper I on the same star observed in two-telescope mode. 
This is especially important because, although the amount of time spent taking data on the star each night can be as long as 
three hours or more, it was more typically
about one hour. Our long set-up procedure prior to science observations often limits our ability to begin data-taking at the start of the night,
and the end point of the observation sequence can be cut short by factors such as (1) the daytime obligations of observing 
team members, (2) losing the star as it transits the meridian 
due to the difficulty of manually guiding near the zenith, where the tracking of our altitude-azimuth 
telescopes is poor, and (3) the maritime climate in New Haven, where high fog 
and patchy clouds can sometimes appear unpredictably.

\begin{figure}[!t]
\figurenum{3}
\hspace{5.5cm}
\includegraphics[scale=0.4]{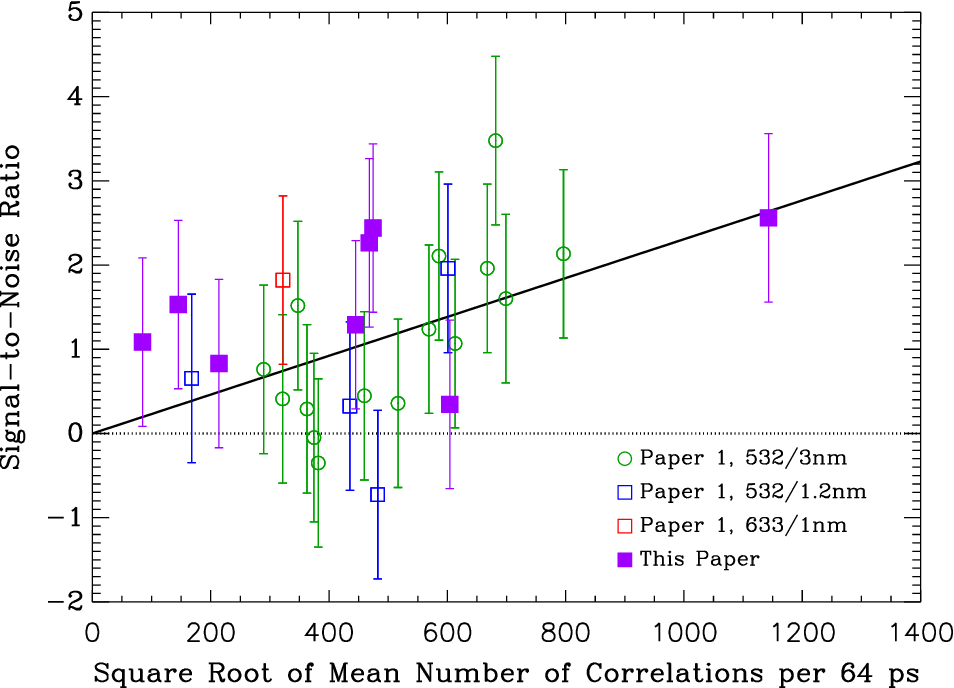}
\figcaption{Signal-to-noise ratio obtained in the final cross-correlation function in each observing session listed in Table 1 as a 
function of the square root of the mean number of correlations obtained in a 64-ps timing bin. 
The filled purple squares indicate the results for the new observations presented in Table 1, all of which were with the 532/1.2 nm filters. 
For comparison, open symbols indicate data from Paper 1, where the filter parameters are given in the legend.
A line of best fit for all data to date is shown as the solid line, 
and the dotted line at zero is present to guide 
the eye.}
\end{figure}

\begin{figure}
\figurenum{4}
\plottwo{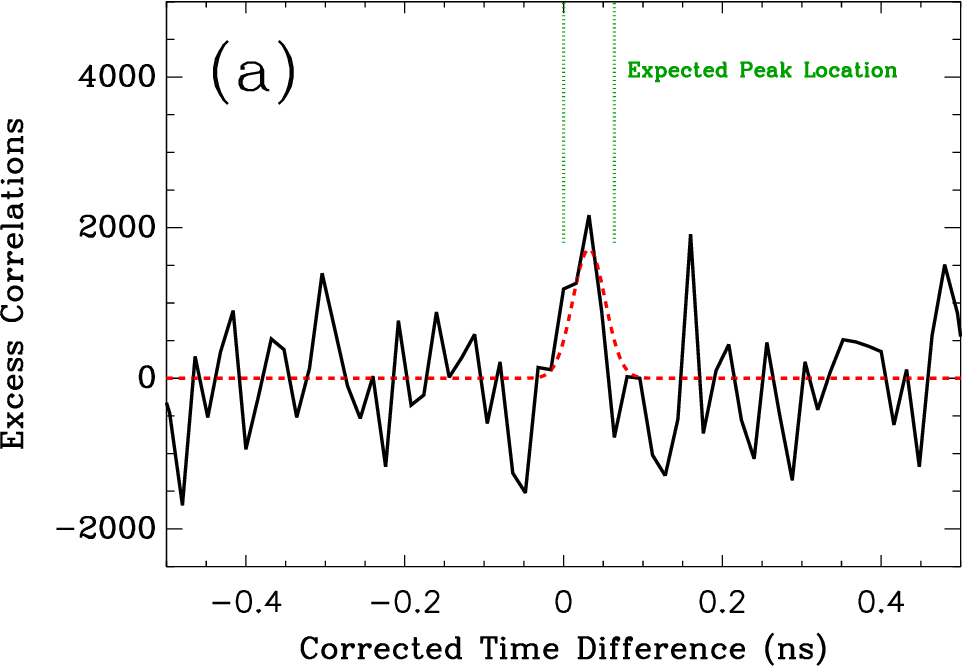}{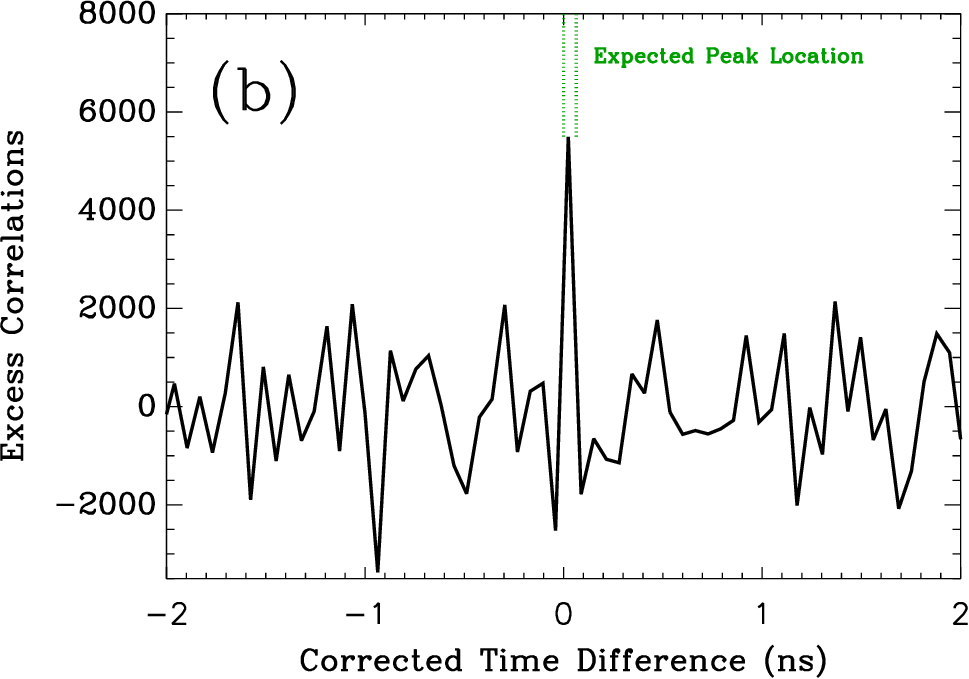}
\figcaption{Cross-correlation results obtained on Vega for the entire new data set presented here. (a) In this case, the plots are made
binning the timing differences between photon pairs in 16-ps bins. The black solid curve is the data, the red dashed curve
is a Gaussian fit to the data, and the green dotted lines indicate the expected location of the peak based on timing
corrections made throughout the dataset. (b) The same data, but using a bin width of 64 ps.}
\end{figure}

\begin{figure}[!t]
\figurenum{5}
\plottwo{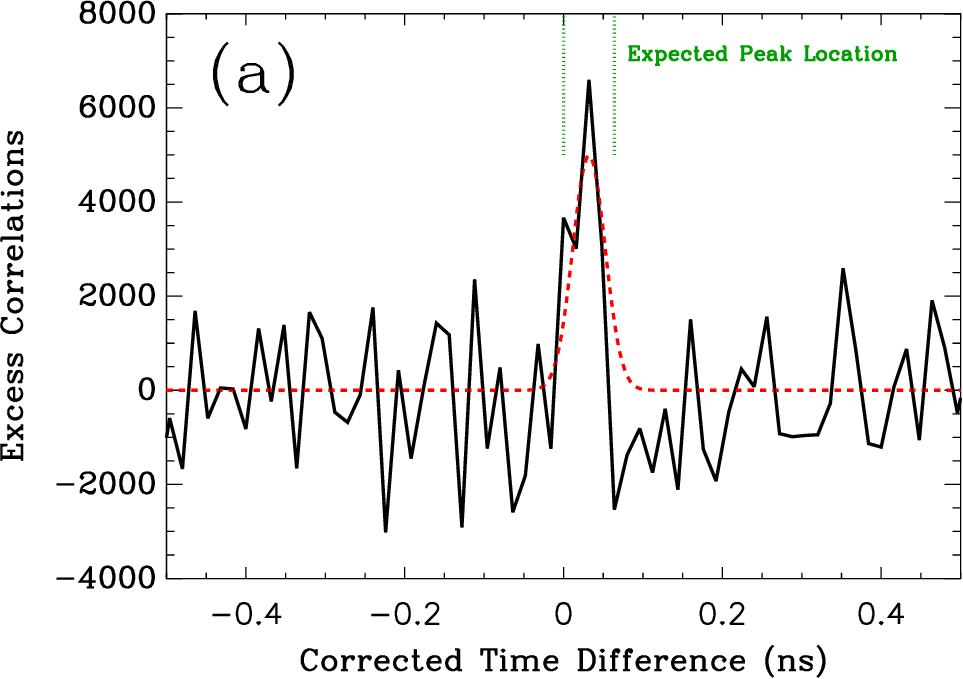}{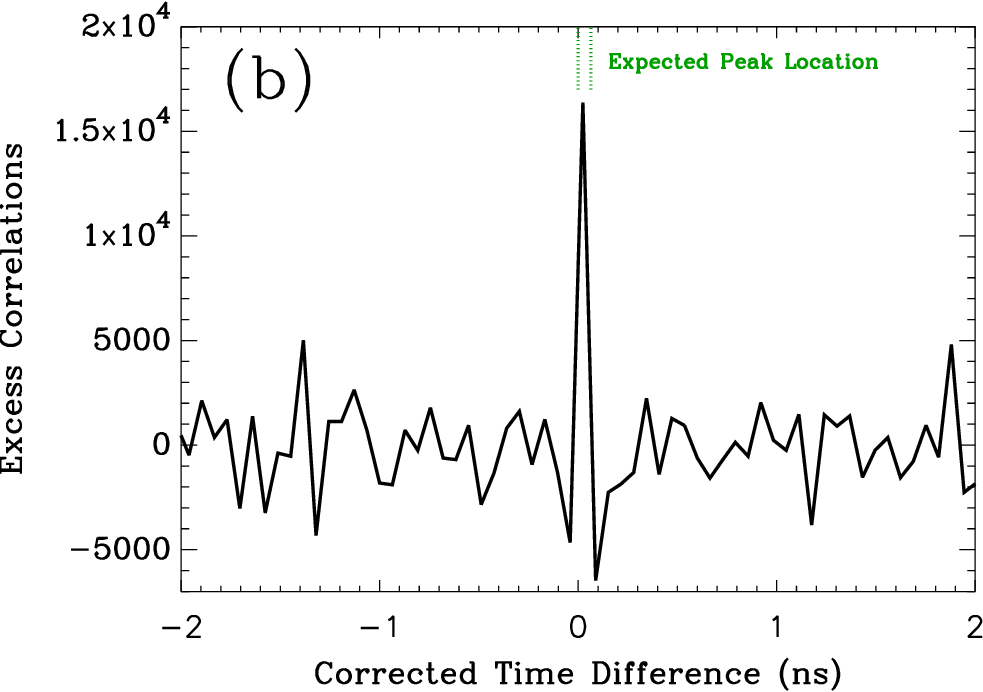}
\figcaption{Cross-correlation results obtained on unresolved sources with SCSI to date by combining the results in Figure 4
with those from Paper I. (a) In this case, the plots are made
binning the timing differences between photon pairs in 16-ps bins. The black solid curve is the data, the red dashed curve
is a Gaussian fit to the data, and the green dotted lines indicate the expected location of the peak based on timing
corrections made throughout the dataset. (b) The same data, but using a bin width of 64 ps.}
\end{figure}

Figure 2 shows the final cross-correlations obtained for each of the eight nights in Table 1. The initial timing differences between 
stations based on the starting sky position of the star were subtracted, and then the timing differences throughout the observing
session were referenced to that fiducial, so that the cross-correlation peak in each case is expected to be in the center of the timing
bin at zero nanoseconds. While it is more typical to plot a version of the HBT peak from 
cross-correlation data that is normalized by the mean number of correlations per timing bin, the eight plots here are shown on the same 
``raw'' scale of
excess correlations, that is, the number of correlations detected in each timing bin after subtracting the average number.
This highlights the variation in observing efficiency, 
with some nights producing better results than others. In the current analysis, we found this useful for comparing to and
combining with
results in Paper I, but for future 
science observations, we plan to adopt the standard normalization as other groups have.
Figure 3 shows the signal-to-noise ratio (SNR) obtained as a function of the square root of the mean number of correlations per 
timing bin. We 
expect that, at constant count rate, the mean number of correlations will be proportional to $t$, the observation time. Since
the SNR will scale as $\sqrt{t}$, the plot of SNR as a function of the square root of the mean number of correlations
is expected to be linear. Fitting a line with intercept zero to the eight points representing the new data on Vega presented here, we
obtain a slope of $m = 0.00269 \pm 0.00065$, which is slightly higher than the value obtained in Paper I, although still within
the uncertainty.




Figure 4 shows the final result for Vega in two ways. In Figure 4(a), 
the timing difference data between channels are binned in 16-ps intervals, and the data show a peak in the expected location.
We fitted this peak with a Gaussian function where the height and location in time were allowed to vary, but the width was fixed. 
As discussed in Paper I, we expect the FWHM in timing resolution for our system to be 48 ps, given the properties
of our SPAD detectors, and so this is the 
fixed value
used in the Gaussian fitting procedure. The result is shown with the red dashed curve in the figure; the height is at 
an excess correlation value of $1821 \pm 484$ and timing offset of $23 \pm 8$ ps. Integrating the Gaussian, a total number of excess 
correlations of $5815 \pm 1544$ is obtained. 
In the case of 64-ps binning, a delta-function peak is expected at the final corrected timing difference between the channels
because the timing jitter of the detectors is lower than that width. In Figure 4(b), a total of 5495 excess correlations were seen
from the summed data set described in Table 1, whereas the noise ({\it i.e.,\ }the standard deviation) of values to the left and right 
of the peak in Figure 3(b) is 1163. This is a formal significance of 4.73$\sigma$. Given that some of the average count 
rates shown in Table 1 are clearly lower than what can be detected with the system under good conditions, further 
refinements in our observing procedures to reliably maximize count rates will likely continue to help us improve our ability to reach 
statistically significant results in a shorter amount of time.

Finally, we combine this result with that shown in Paper I, in order to update our instrumental efficiency for unresolved
sources. It is important to obtain as precise a number as possible for the expected correlations when a source is unresolved prior
to interpreting data from a partially-resolved star as we will do in the next section. The results are shown in Figure 5 for both the 
16- and 64-ps binning. For 16-ps binning, we again fit with a Gaussian peak of fixed FWHM at 48 ps, and we find a peak value of 
$5193 \pm 595$ correlations, and the peak position sits at $25 \pm 3$ ps from the origin. Integrating the Gaussian, we obtain
the total number of excess correlations is $16582 \pm 1901$, a formal significance of 8.72$\sigma$.
For 64-ps binning, we obtain 
16375 excess correlations with a noise value of 1797 on either side of the peak, a formal significance of 9.11$\sigma$. As this 
result is obtained with mean number of correlations of $7.478 \times 10^{6}$ per 64-ps bin, the implied correlation efficiency of the 
instrument is $0.2190 \pm 0.0240$\%. This is the value we will use in the next section to represent SCSI's HBT signal 
for an unresolved source.




\section{Observations of Arcturus}

\subsection{Observations with SCSI}

We now turn to our observations of Arcturus (= HR 5340 = HIP 69673), where in all cases we observed the star in two-telescope
mode using a north-south baseline.
In Paper I, observations of Arcturus that were obtained on four nights were presented, but they represented only 3.35 hours of 
data. While those data
show that the level of correlation detected was lower than that of Vega and Deneb (which were both unresolved at the
baselines used), it was not possible to use the measurement to conclude anything other than a rough
lower limit on the diameter
of the star. In the first half of 2025, we sought to continue observing this star with SCSI to decrease the uncertainty of the previous 
measure. The list of these observations is presented in Table 2. We also were able to re-analyze one night of Arcturus data
that occurred on 2017 June 21 and was not presented in Paper I; this was carried out before we obtained the narrower 
($\Delta \lambda = 1.2$ nm)
filters that are currently used in SCSI; nonetheless, we include it here for completeness. The dataset represents $\sim$8 hours of
observations at an average count rate of $\sim$1 MHz. Combined with the earlier dataset from Paper I, 11 hours of data on the
star have now been recorded, comparable to the new dataset presented of Vega in the previous section.

We present the data in the same way as for Vega. The cross-correlation functions obtained on each night of new data are shown in
Figure 6, and the SNR obtained in each case is shown in Figure 7 as a function of the square root of the 
mean number of correlations obtained per 64-ps
timing bin. In this plot, we also include points for the 4 nights reported in Paper I. Fitting all of the data with a linear function
with zero intercept, we obtain a slope of  $m = 0.00118 \pm 0.00070$,
or about half of the value obtained for Vega discussed in the previous section, suggesting partial correlation. 
Figure 8 shows the summed cross-correlation plot for the
new data, at both (a) 16-ps and (b) 64-ps resolution. In the former case, a Gaussian of 48 ps is again fitted to the data, and the
peak position falls closer to the high edge of the expected range in timing difference (that is, at $\sim$50 ps instead of 
exactly in the middle of the 0-63 ps interval). One difference from the 
Vega data presented earlier and also two-telescope data from Paper I is that we most often used Telescope 3 in place of
Telescope 2 as the northern station. Thus, if the focal length of Telescope 3 is slightly different from Telescope 2, changing
the optical path length by a few millimeters, this could explain such a shift.
Figure 9 is the same as Figure 8, but includes the Arcturus data from Paper I in the summations (and we make no correction
for the slight offset just mentioned).
These plots appear very similar to Figure 8 due to the fact that the combined dataset is dominated by the new observations; 
nonetheless, we include
the older data as it provides an incremental increase in the SNR overall.

\begin{deluxetable}{ccccccc}
\tablewidth{0pt}
\tablenum{2}
\tablecaption{New SCSI Observations of Arcturus}
\tablehead{
\colhead{Date} & 
\colhead{N-S Sep.} &
\colhead{Total Observing} &  \colhead{Avg. Count Rate}\\
(UT) & (m) &  Time (hr)\tablenotemark{a} & per Tel. (MHz)
}
\startdata
2017 Jun 21 \tablenotemark{b} & 2.331 & 0.33 & 0.272 \\
2025 Apr 18 & 3.005 & 0.48 & 1.333\\
2025 Apr 25 & 2.359 & 1.27 & 1.197\\
2025 May 01 & 2.129 & 1.25 & 1.314 \\
2025 May 12 & 2.151 & 1.11 & 0.770 \\
2025 May 27 & 2.224 & 1.75 & 1.165 \\
2025 Jun 04 & 2.061 & 1.58 & 0.546 \\
\enddata
\tablenotetext{a}{There are 7.77 total hours of data obtained on the seven nights represented here.}
\tablenotetext{b}{This observation taken with the $\lambda/\Delta \lambda = 532/3.0$ filter.}
\vspace{-0.5cm}
\end{deluxetable}

\begin{figure}[!t]
\figurenum{6}
\plottwo{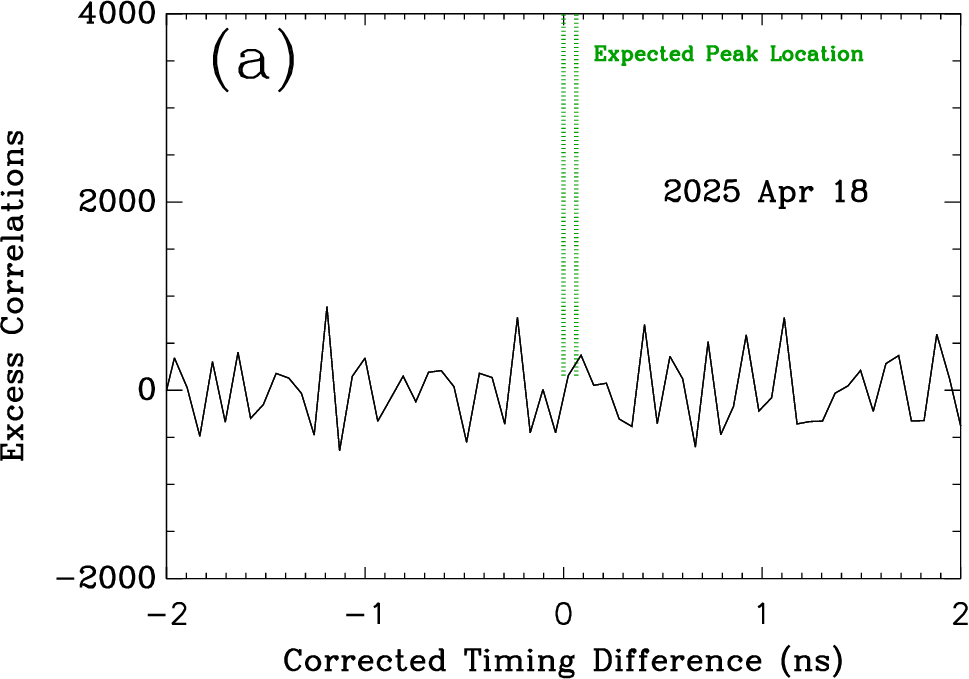}{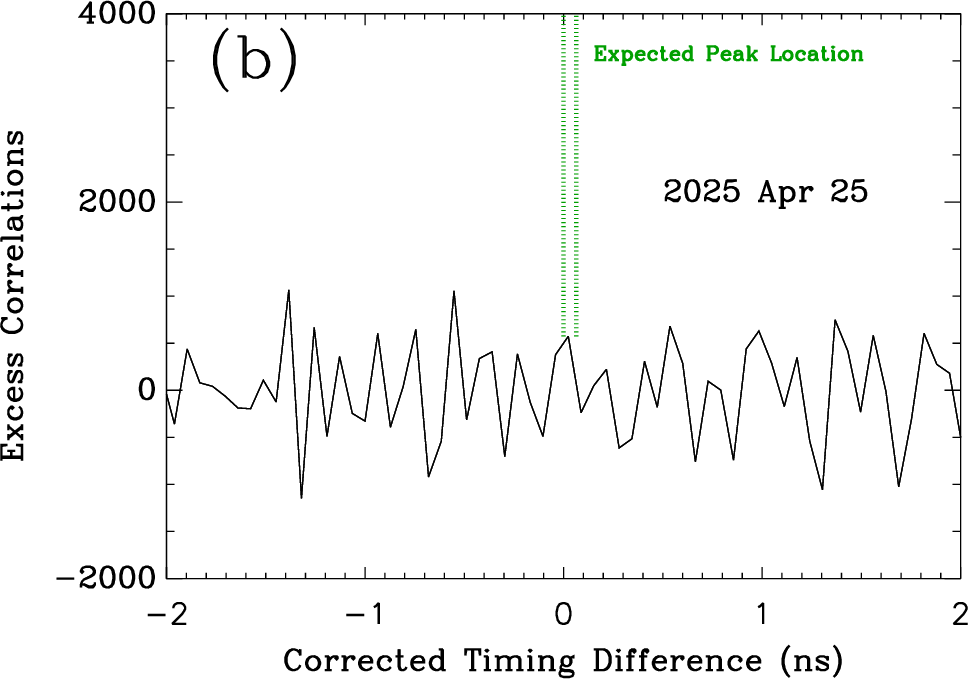}
\plottwo{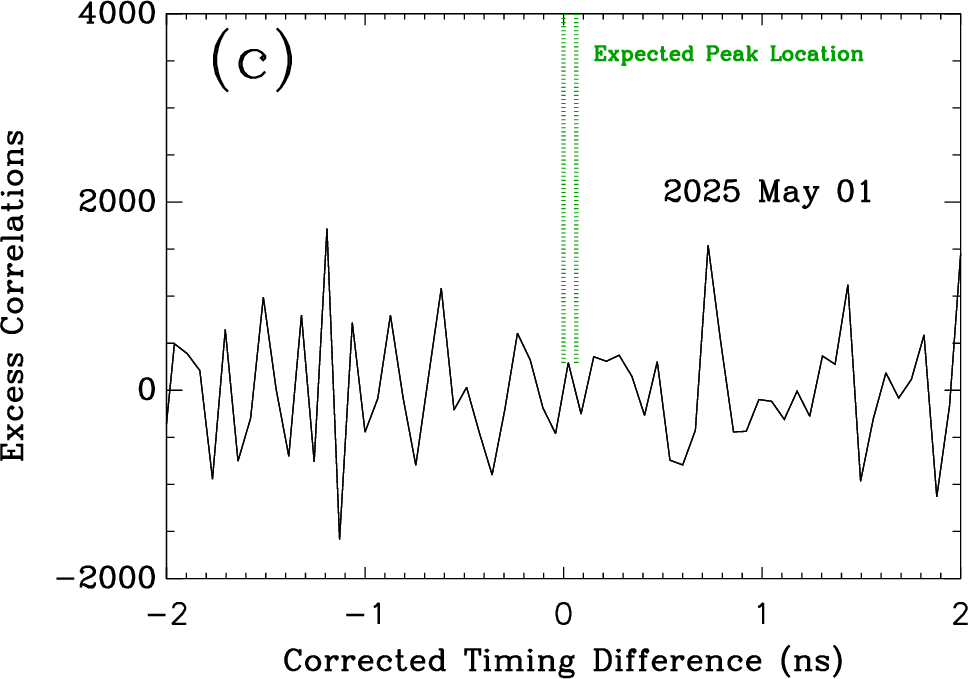}{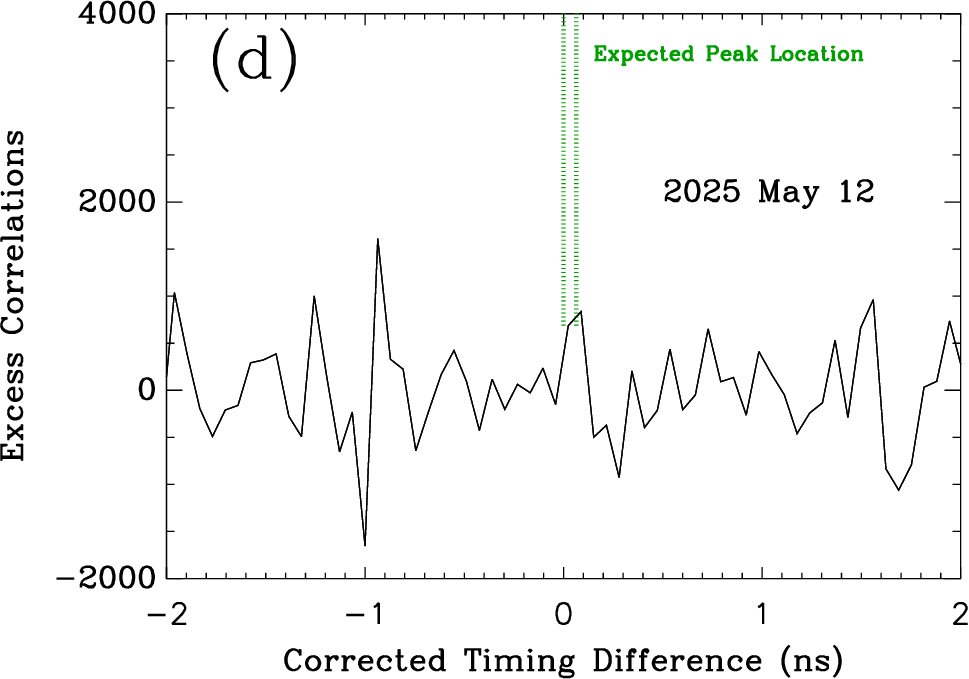}
\plottwo{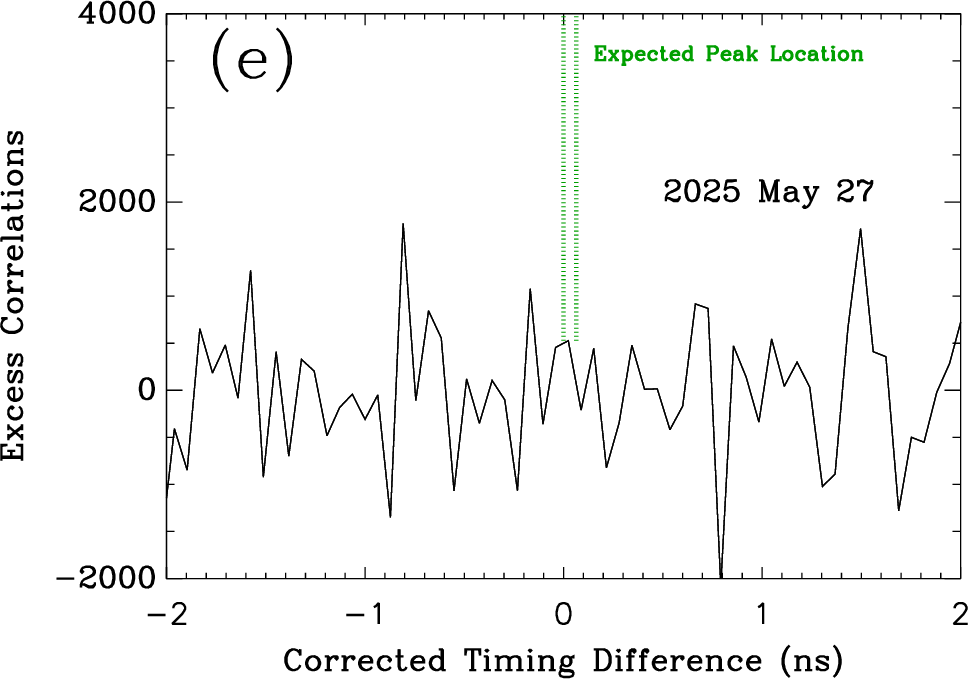}{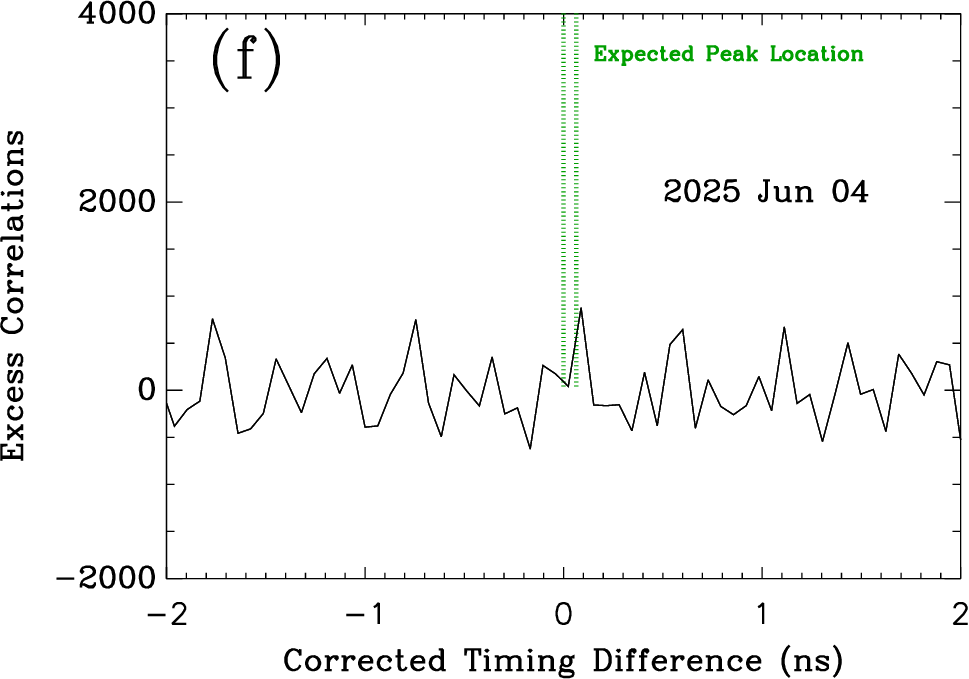}
\includegraphics[scale=0.5]{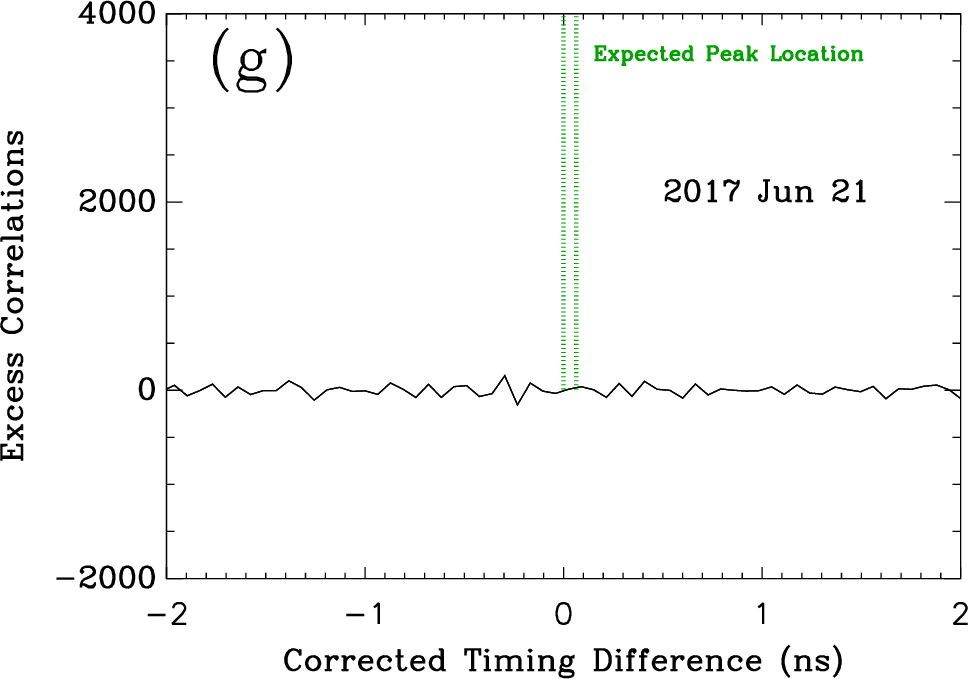}
\figcaption{Cross-correlations obtained for each night listed in Table 2. In each case, the mean number of correlations
per timing bin has been subtracted so that the HBT peak is more visible, and the green dotted lines mark the 
expected location of the peak. The two-telescope arrangement discussed in the text was used for all observations here.}
\end{figure}

\begin{figure}[!tb]
\figurenum{7}
\hspace{5.5cm}
\includegraphics[scale=0.4]{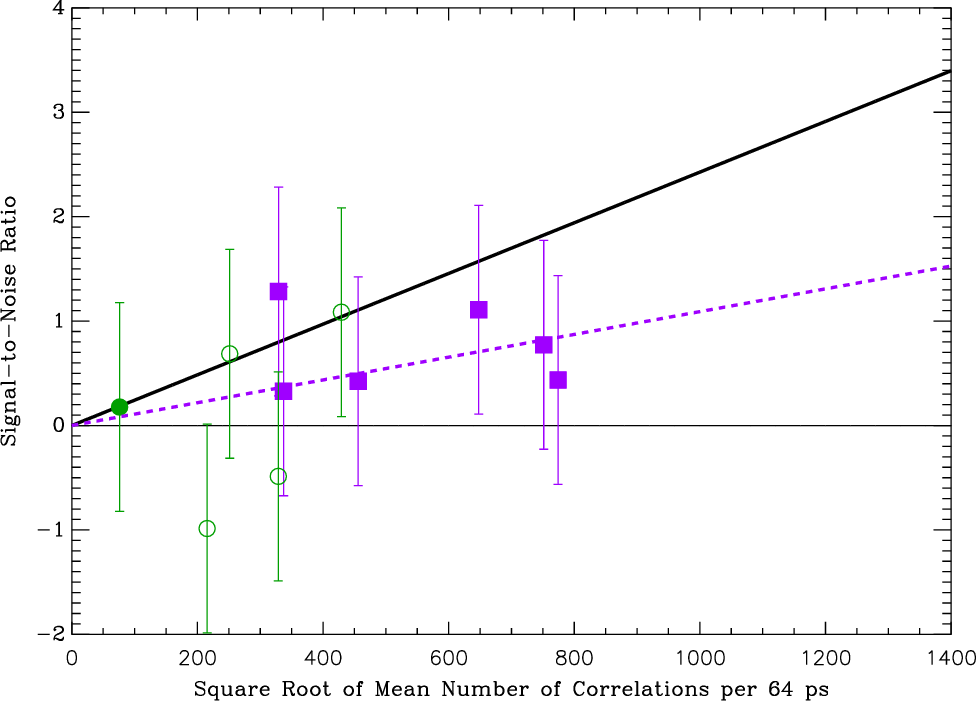}
\figcaption{Signal-to-noise ratio obtained in the final cross-correlation function in each observing session listed in Table 2 
as a function of the square root of the mean number of correlations obtained in a 64-ps timing bin. 
The purple squares indicate the results from Table 2, except for the 2017 Jun 21 data, which is shown as a solid green
circle. For comparison, 
we also include here the data from Paper I as open green circles. Both the 2017 Jun 21 data and the Paper I data were
obtained with a filter of center wavelength/FWHM of $\lambda$/$\Delta \lambda$ = 532/3nm. 
The line of best fit from Figure 3 is shown
as the solid line, the dashed purple line indicates the line of best fit for Arcturus, and the solid line at zero is present to guide 
the eye.}
\end{figure}

\begin{figure}
\figurenum{8}
\hspace{1.3cm}
\includegraphics[scale=0.43]{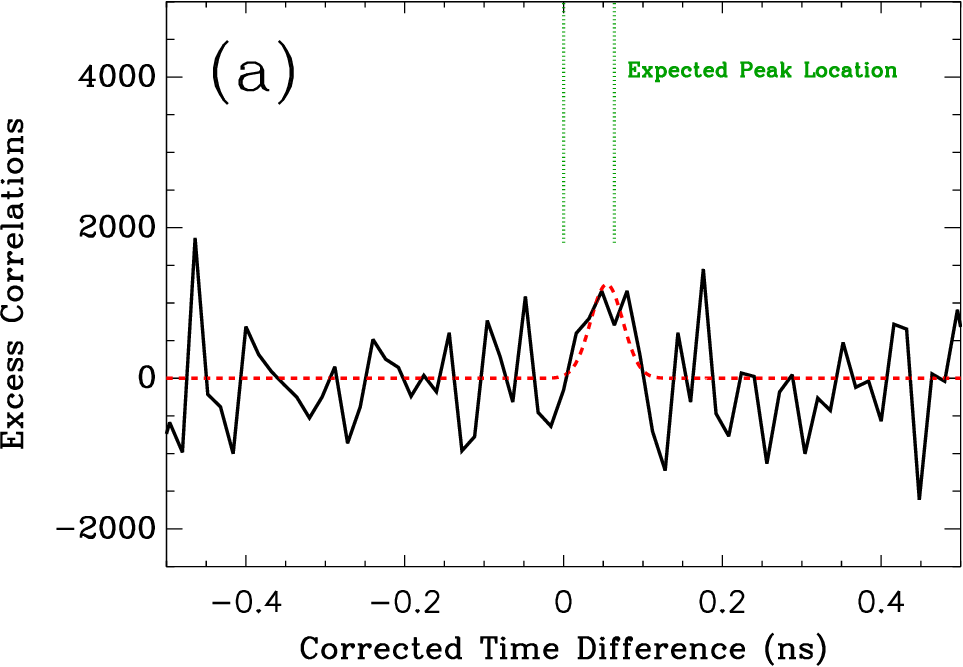}
\hspace{1.0cm}
\includegraphics[scale=0.43]{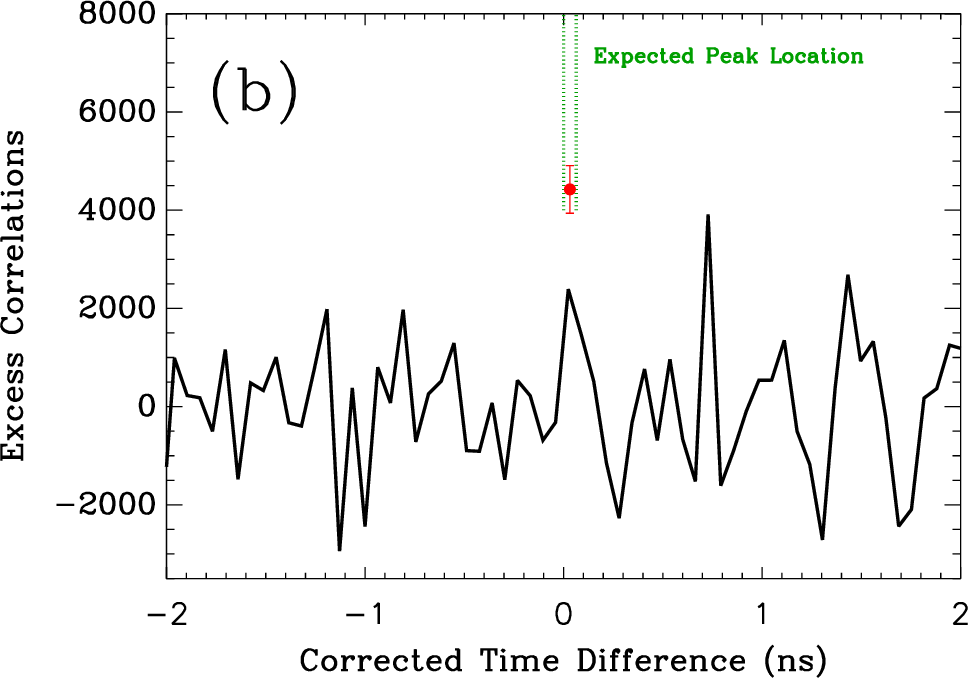}
\figcaption{Cross-correlation results obtained on Arcturus for the new data set presented here. (a) In this case, the plots are made
binning the timing differences between photon pairs in 16-ps bins. The black solid curve is the data, the red dashed curve
is a Gaussian fit to the data, and the green dotted lines indicate the expected location of the peak based on timing
corrections made throughout the dataset. (b) The same data, but using a bin width of 64 ps. The red filled circle indicates
the signal value expected for full correlation based on the results from Vega in the previous section.}
\end{figure}

\begin{figure}[!b]
\figurenum{9}
\hspace{1.3cm}
\includegraphics[scale=0.43]{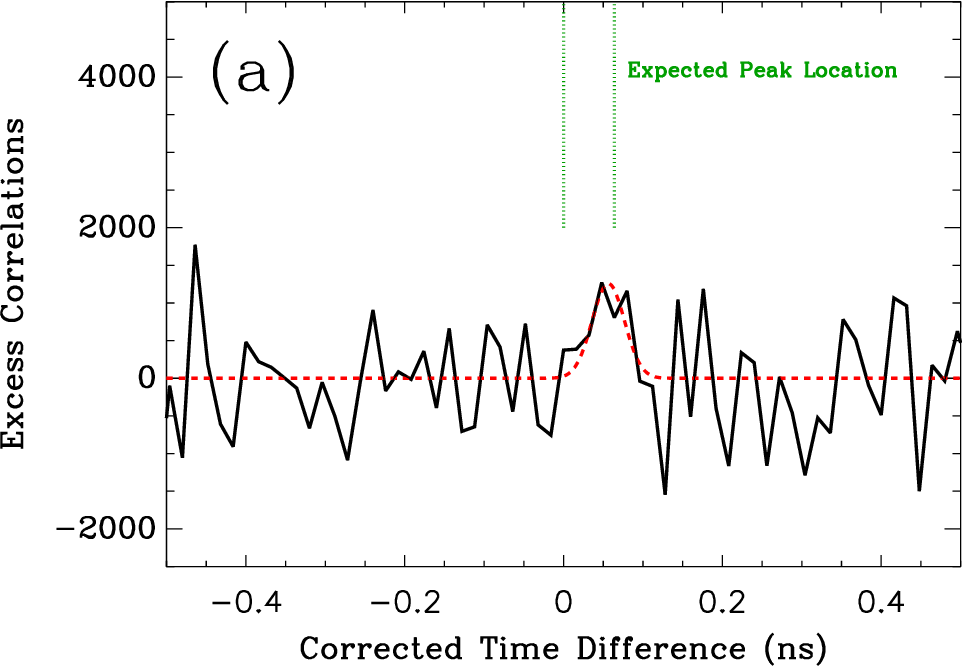}
\hspace{1.0cm}
\includegraphics[scale=0.43]{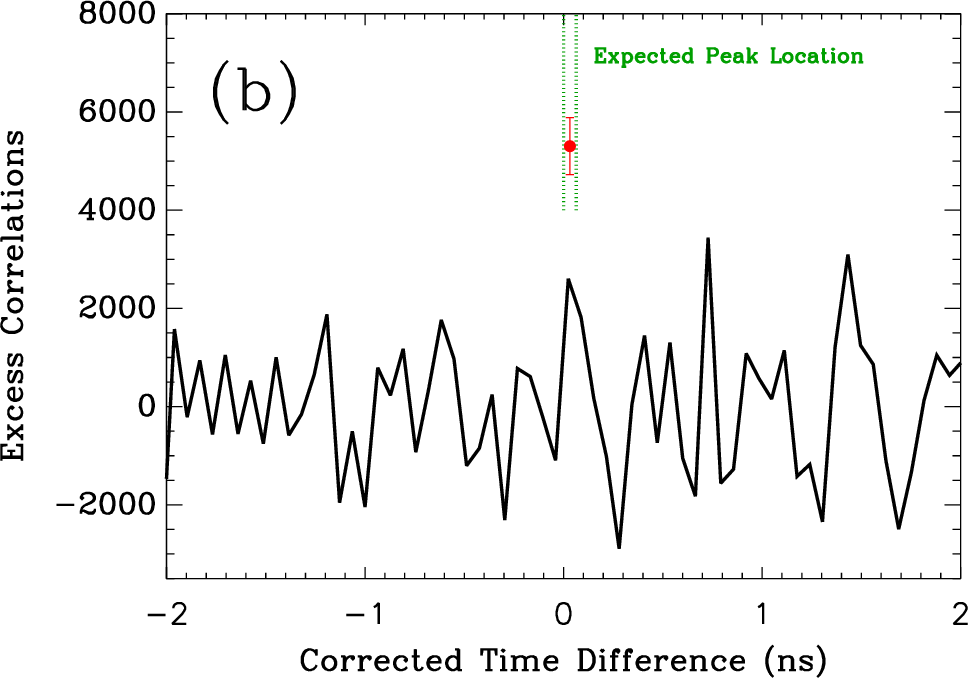}
\figcaption{Cross-correlation results obtained on Arcturus for the data set presented here plus the results in 
Paper I. (a) In this case, the plots are made
binning the timing differences between photon pairs in 16-ps bins. The black solid curve is the data, the red dashed curve
is a Gaussian fit to the data, and the green dotted lines indicate the expected location of the peak based on timing
corrections made throughout the dataset. (b) The same data, but using a bin width of 64 ps. The red filled circle indicates
the signal value expected for full correlation based on the results from Vega in the previous section.}
\end{figure}

For Figure 8, we obtain a peak value at 16-ps binning that has value  
$1247 \pm 454$, 
with
the position of the peak at $54.4 \pm 10.5$ ps. The integral of the fit, representing the total number of excess correlations is
$3980 \pm 1449$. 
For the 64-ps binning shown in Figure 8(b), we have a measured signal in the expected timing bin
of 2390 counts, with the standard deviation of values on either side of this location being 1321. This implies a SNR 
of 1.810, with the mean number of correlations per 64-ps bin being $2.02 \times 10^{6}$
counts.

In Figure 9, we combine these results with those presented in Paper I. Here, for 16-ps binning and fitting in the same way as
for Vega, we obtain a peak value of 
$1258 \pm 476$ 
at $56.3 \pm 10.9$ ps. The integral of the fit implies a total of 
$4018 \pm 1520$ 
excess counts. For 64-ps binning, we obtain an excess of 2603 counts in the expected timing bin, with 
a standard deviation of 1370 counts. Thus, the SNR here is 1.90 after having obtained 
$2.42 \times 10^{6}$
correlations per 64-ps
bin in the summation of all files. In both the case of Vega and Arcturus, the results of binning at 16 ps show that we do not
miss a significant number of counts by binning the correlations by 64 ps, nor do we increase the noise. It is therefore reasonable
in what follows here to use only the results binned at 64 ps. More sophisticated analyses using higher timing resolution
may be possible, but it is unlikely that 
they would dramatically increase the statistical significance of the results over this simple and straightforward approach.

The data presented allow us to estimate the squared visibility of Arcturus for the baseline of our observations. Dividing the
excess correlations by the mean number of correlations per 64-ps bin, we obtain $0.107 \pm 0.056$\%, which represents
the percentage of intensity-correlated photons detected. For unpolarized 
light, the relationship between the cross-correlation results and the complex visibility function of the source is expected to be
\begin{equation}
 g^{(2)}_{\rm unpol} = 1 + \frac{1}{2} |V_{12}|^2 \frac{\Delta \tau}{\Delta t},
\end{equation}
where $g^{(2)}_{\rm unpol}$ is the cross-correlation function in the case where the mean number of correlations per timing bin 
is normalized to 1, $ |V_{12}|^2 $ is the squared visibility, $\Delta \tau$ is the 
timescale of intensity fluctuations set by the frequency bandwidth of the system, and $\Delta t$ is the timing resolution used.
The measurements of excess correlation discussed here effectively subtract the 1 on the right side of the equation
and are represented by the second term on the right; thus, to obtain
the visibility measurement for Arcturus, we divide the correlation percentage obtained for Arcturus by that of Vega from the previous
section. The final result is $ |V_{12}|^2 = 0.49 \pm 0.26$. The range of spatial frequencies to which this result applies will be 
discussed in Section 4.


\subsection{Speckle Imaging of Arcturus}

In June of 2025, three of us (E.P.H., J.W.D., and S.R.M.) had the opportunity to observe Arcturus using speckle imaging
at the ARC 3.5-m Telescope located at Apache Point Observatory (APO) in New Mexico. The instrument used for these 
observations was the Differential Speckle Survey Instrument (DSSI) \citep{2009AJ....137.5057H}. As detailed in 
\citet{2023AAS...24130506H} and \citet{2024AJ....167..117D}, 
DSSI was moved from its previous home at Lowell Observatory to APO in early 2022, and since that time Southern 
Connecticut State University has 
collaborated with the University of Virginia to build a long-term speckle program there. 

Arcturus was observed on two nights, 08 and 11 June 2025 UT. In both cases, 3000 1-ms frames were taken of the star, 
followed by 3000 1-ms frames taken of a bright unresolved star, HR 5405 (HD 126661, 22 Boo). HR 5405  is 
a late A or early F giant \citep{2001AJ....121.2148G} at a distance of 96 pc, so its inferred diameter
based on stellar properties is $4.03 \pm 0.25 R_\odot$ \citep{2018AA...616A...1G}, and therefore the angular diameter is
 approximately 0.39 mas, which is sufficiently small to be considered as completely unresolved for our purposes, 
 given that Arcturus is partially resolved over a baseline
 of 2-3 meters, as shown in the previous section.
DSSI collects data in two filters simultaneously; in 
our case, those filters were a red ($\lambda_0 = 692$ nm, $\Delta \lambda = 40$ nm) and near infrared ($\lambda_0 = 880$ nm, 
$\Delta \lambda = 50$ nm). The seeing measured on both nights was between 0.8 and 1 arcsecond.

The reduction of the speckle frames follows the standard autocorrelation analysis that has been described in e.g.\
\citet{2024AJ....167..117D} and other papers. The basic process is to debias frames and compute their autocorrelations,
and then sum the autocorrelations for the entire sequence. This retains diffraction-limited information. Completing the same steps for the 
point source observations, we can then deconvolve the speckle transfer function from the data on Arcturus by moving to 
the Fourier domain and dividing. A comparison of the Fourier transform of both Arcturus' autocorrelation (which is equivalent
to its spatial frequency power spectrum) and that of the point source is shown for each filter in Figure 10. It is seen that 
at higher spatial frequencies, there is a small but statistically significant loss of power in the Arcturus data relative to that 
of HR 5405, indicating that, while Arcturus' angular diameter is below the diffraction limit at the ARC telescope, there is 
a partial resolution of the star at the highest spatial frequencies represented.

\begin{figure}[!t]
\figurenum{10}
\plottwo{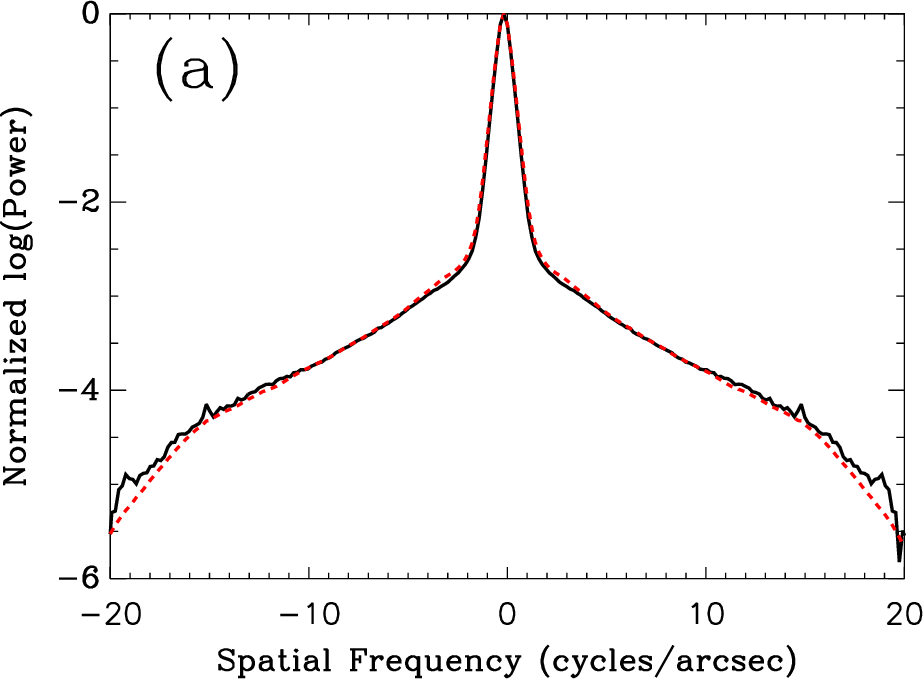}{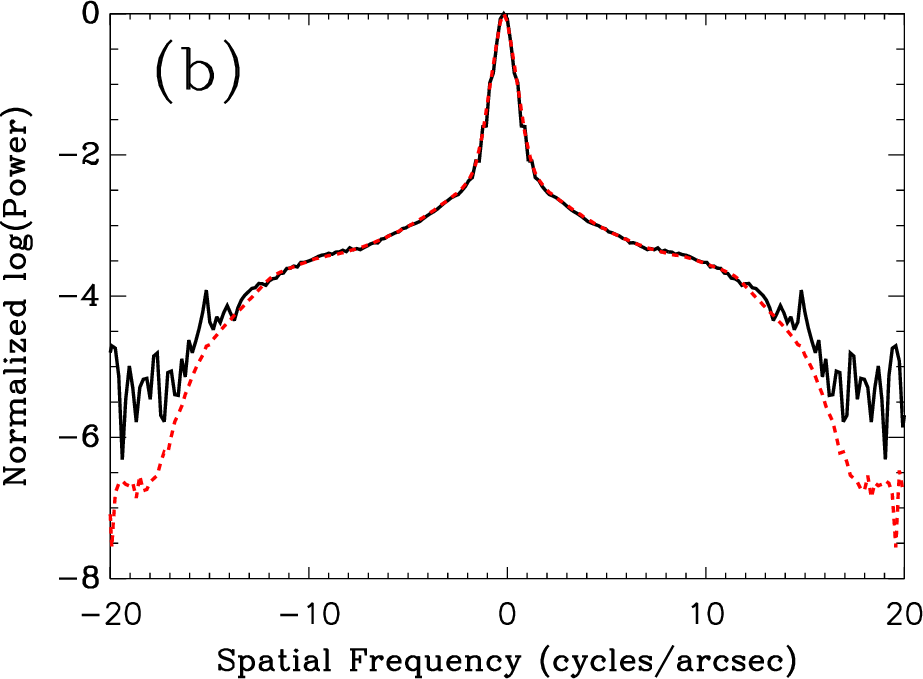}
\figcaption{A cross section of the two-dimensional power spectrum obtained using speckle imaging. (a) The result at 
692 nm. (b) The result at 880 nm. In both plots, the solid line is the power spectrum of the unresolved star, HR 5405, and
the red dashed curve is the power spectrum of Arcturus. The large peak at low frequencies is caused by seeing, and 
in both cases, the curves are nearly identical at lower spatial frequencies, but begin to slightly diverge at the higher
frequencies shown, with Arcturus having lower power in that region. Note that the curves do not 
extend to as high a limit for the 880-nm data because the diffraction limit is reached at a lower spatial frequency for the
longer wavelength. A higher noise floor is also seen above the diffraction limit for the point source at 880-nm because it 
is a fainter star than Arcturus.}
\end{figure}

\section{A New Measurement of the Diameter of Arcturus}

By combining the data from both SCSI and DSSI, we are in a position to measure the visibility function of 
Arcturus. The spatial frequency power spectrum of the star, once deconvolved by that of the point source HR 5405
and normalized, yields the 2-D squared visibility at low spatial frequencies. The intensity interferometry data 
supplement our information at higher spatial frequencies. To help visualize the information we have, Figure 11(a)
shows the spatial frequency coverage of our data, where we have considered only spatial frequencies up to 16
cycles/arcsec for the speckle imaging data in order to maintain high signal-to-noise results. In Figure 11(b), 
we radially average the data to obtain a visibility curve as a function of spatial frequency. In the case of the speckle 
data, we use annuli of width 2 pixels in the frequency domain centered on spatial frequencies of 2, 4, 6, ..., and 18 
cycles per arcsecond for 692-nm speckle data, and 2, 4, 6, ..., and 12 cycles per arcsecond for the 880-nm speckle data. 
We do this for each filter on both nights, generating four independent estimates of the squared 
visibility for these frequencies. Individual results of each observing session with SCSI are then also plotted, along
with the average value presented in the previous section.

\begin{figure}[!t]
\figurenum{11}
\includegraphics[scale=0.65]{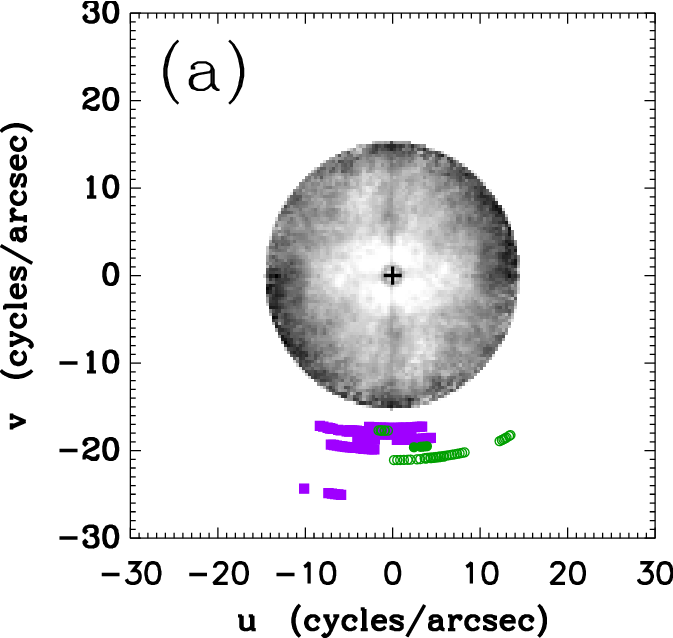}
\hspace{0.5cm}
\includegraphics[scale=0.62]{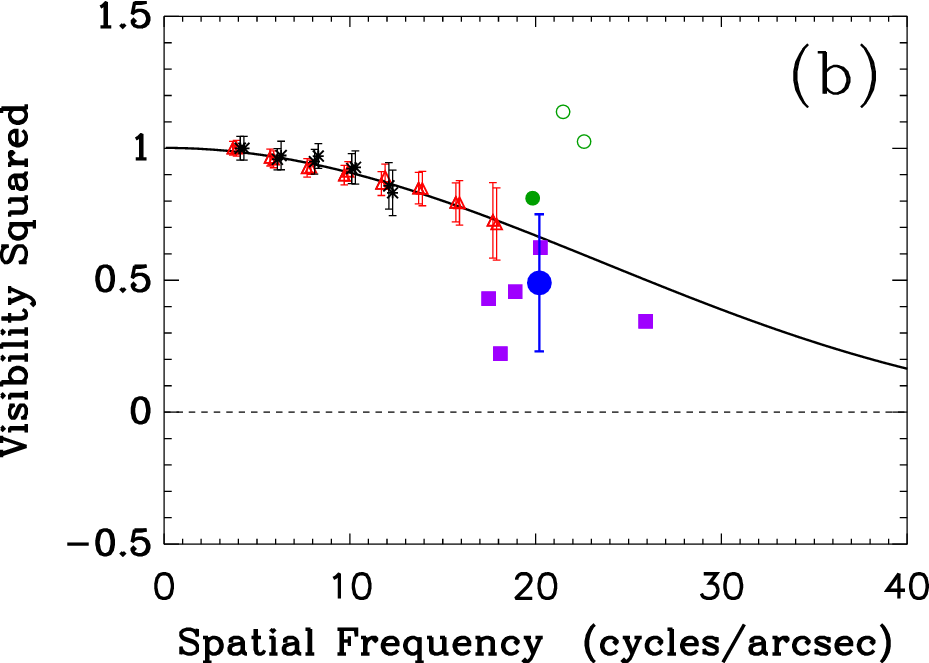}
\figcaption{(a) The spatial frequency coverage for Arcturus. The grayscale image inside the circle represents the derived 
spatial frequency
power spectrum for the averaged speckle data, extending out to 15 cycles/arcsec, with darker shades near the 
edge of the circle indicating lower power.
The cross marks the origin in the $uv$-plane,
and the purple and green data points illustrate the observing sessions with the intensity interferometer. The color coding 
and symbols are the same as in Figure 7. (b) The visibility curve for Arcturus. Here, the red triangles are the radially
averaged values of visibility for the 692-nm speckle data, the asterisk symbols are the 880-nm speckle data, and the
blue filled circle is the summed result for the intensity interferometry data. Green and purple squares indicate the visibilities
obtained for individual observing sessions, albeit with large uncertainty in each case. The colors and symbols are the 
same as in Figure 7, and three observations included in the final visibility measure fall outside the plot window. 
The solid curve is the fitted visibility as a function of spatial frequency, and the dashed curve is
drawn at zero to guide the eye.}
\end{figure}

Figure 11 also plots the best fit to the data for a radially-symmetric limb-darkened profile. 
As discussed in \citet{2018AJ....155...30B}, if $x = \pi B \theta_{LD} \lambda^{-1}$ where $B$ is the baseline, 
$\theta_{LD}$ is the limb-darkened stellar diameter, and $\lambda$ is the observation wavelength, then one may model the 
visbility squared for a limb-darkened disk as

\begin{equation}
V^2 = \left( \frac{1-\mu}{2} +\frac {\mu}{3} \right)^{-1} \times 
\left[ (1-\mu)\frac{J_{1}(x)}{x} + \mu \left( \frac{\pi}{2} \right)^{1/2}\frac{J_{3/2}(x)}{x^{3/2}} \right]^{2},
\end{equation}

\noindent
where $\mu$ is the limb darkening parameter, which is generally wavelength dependent. In our case, we have data at three different
wavelengths, 692 nm and 880 nm from the speckle observations and 532 nm from the intensity interferometry observations. Baines
et al.\ derived values for $\mu$ for a range of stellar temperatures and assuming observing in the $R$ band; generally, for stars with 
similar temperatures as Arcturus, the value for $\mu$ was $\sim0.8$. As our highest signal-to-noise data (692 nm) are also near the 
$R$ band, with some data above and below that value, we adopt $\mu = 0.8$ for the analysis here. In principle, with more data, 
we could fit for $\mu$ at each wavelength, but in this study we simply fit the data to the 
function in Equation 2 to a two-parameter fit: $\theta_{LD}$
and an overall (arbitrary) normalization constant. To avoid systematic error due to seeing mismatch between Arcturus and the 
point source, we do not consider spatial frequencies below 4 cycles per arcsecond. Likewise, we only use the speckle data up to 
18 cycles/arcsec at 692 nm, and up to 12 cycles per arcsec at 880 nm, since the SNR deteriorates quickly above these values. 
We normalize the values we have for each filter to 1
at 4 cycles per arcsec, and then perform the two--parameter fit. The result of that operation is $\theta_{LD} = 21.5 \pm 1.8$ mas,
with a normalization of $1.001 \pm 0.013$. 

\section{Discussion}

Several other studies exist in the literature which measure the angular diameter of Arcturus. The first direct measure
was due to \citet{1931ErNW...10...84P}, and speckle interferometry 
contributed two measures in the 1970's using the Palomar 5-m telescope and the Kitt Peak 4-m telescope,
respectively \citep{1972ApJ...173L...1G, 1976PASP...88...69W}. While Pease did not give an uncertainty for his measure,
the speckle results were stated with uncertainties of $\sim$14 and $\sim$35\%, respectively.
Starting in the 1980's, a series of high-precision measures have been made using long baseline optical and infrared interferometers.
\citet{1986A&A...166..204D} presented a measurement using the I2T interferometer at CERGA Observatory in France. The observations
were made in three filters centered near 2.2 microns but having different widths, using baselines of 18-34 m.
\citet{1998A&A...331..619P},
\citet{2005A&A...435..289V}, and
\citet{2008A&A...485..561L} present measures using IOTA (Mt. Hopkins, Arizona) that span over a decade, taken in the $K$-band
(2.2 microns) for the first two studies, and then the $H$-band (1.6 microns) for the third result.
A result using the Mark III Optical Interferometer is given in \citet{1996A&A...312..160Q}, while
 \citet{2003AJ....126.2502M} detail a measurement using the Navy Precision Optical Interferometer.
These were both done at several visible wavelengths (between 450 and 800nm) using baselines of less than 10 m.
\citet{2009MNRAS.399..399R} provide a measurement using VINCI at VLTI, where the observations were performed at the 
$K$-band. 
Finally, \citet{2018A&A...620A..23O} used the AMBER instrument at VLTI to spatially resolve
the star in individual CO line wavelengths, finding evidence for a cool component extending out to 2-3 times the radius of the star.

\citet{2011ApJ...743..135R} provided a summary of diameter measurements made of Arcturus from the 1980's through 2011, where 
they corrected measurements for limb-darkening if the original work had not done so, and they derive a final weighted diameter
of $21.06 \pm 0.17$ mas.  \citet{2018A&A...620A..23O} came after that paper, and they gave a uniform
disk diameter of $20.4 \pm 0.2$ mas for the CO continuum at 2.3 microns. Following \citet{2011ApJ...743..135R}, we estimate that
this value underestimates the limb-darkened diameter by approximately 0.5 mas at 2.3 microns, leading to a limb-darkened
value of $20.9 \pm 0.2$ mas.

\begin{figure}[!t]
\figurenum{12}
\hspace{2.5cm}
\includegraphics[scale=0.8]{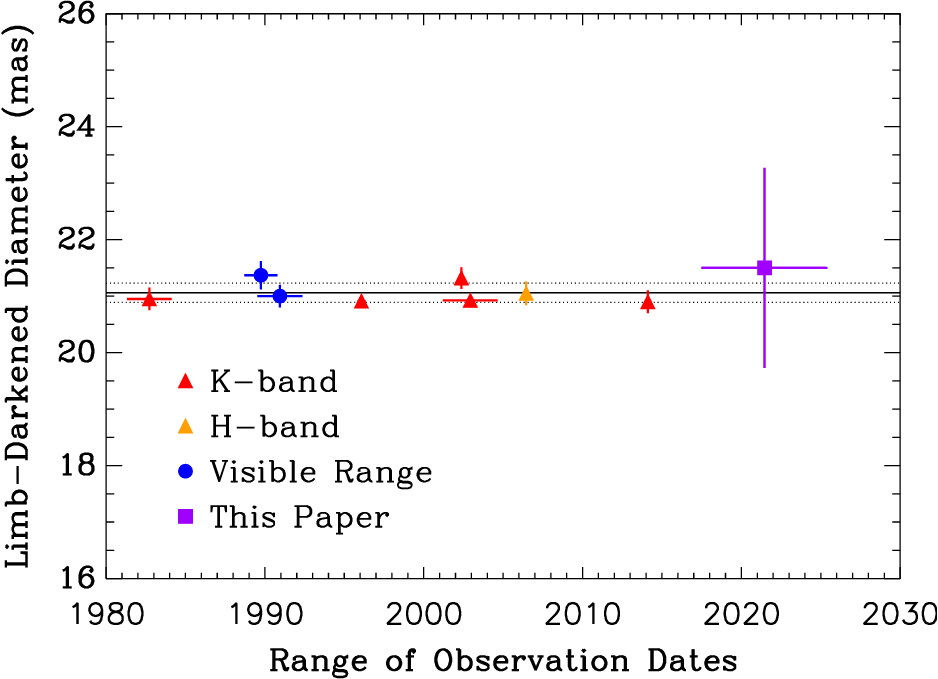}
\figcaption{Diameter measurements of Arcturus made by various groups since 1980 using interferometric techniques. 
$K$-band measures are from I2T, IOTA, and VLTI, the $H$-band data point is from IOTA, the visible measures are made
by the Mark III interferometer and NPOI, and the purple square indicates the measure presented
here. Vertical error bars are included for all measures, and the horizontal error bars indicate the timeframe over which the 
observations were taken in each case. For our measure, though the timeline extends from our earliest observations in 2017
to the present, most of the data are from 2025.
The black solid and dotted lines indicate the weighted average and uncertainty found in \citet{2011ApJ...743..135R}, $21.06 \pm 0.17$ 
mas.}
\end{figure}

In Figure 12, we show the long baseline optical interferometry measurements mentioned above together with our own measure. 
(That is, we exclude the 1931 result of Pease and the early speckle results.)
While results from the Michelson interferometers show higher precision than the result from SCSI presented here, 
repeating the measurement, especially with a technique in the visible
that differs from visible-light Michelson interferometers like the Mark III or NPOI, is useful. Ironically, most of the
other modern intensity interferometers that have produced diameter measurements of stars
 in the last few years do not have the capability of 
observing at small enough baselines to detect the HBT signal from Arcturus. 
Taken as a whole, the diameter measures in Figure 12 show consistency through the years, 
across a factor of more than 4 in wavelength,
and, now, with our result, between the two 
techniques represented (Michelson interferometry and intensity interferometry). Further observations in 3-telescope mode with 
SCSI will
of course help us reduce the uncertainties in the intensity interferometry result.

\section{Conclusions}

We have presented our second large set of observations using the Southern Connecticut Stellar Interferometer (SCSI). 
In the first group of observations presented here, we used SCSI in a three-telescope mode for the first time. We observed Vega, producing 
a nearly 5-$\sigma$ photon bunching peak in roughly 12 hours on sky. This allowed us to combine with the results from our
first paper to update the fraction of correlated photons seen with our instrument to $0.2190 \pm 0.0240$\%. In the spring of 
2025, we were then able to observe Arcturus on 6 nights, finding a reduced fraction of correlated photons in that case.
Forming the ratio of these two correlation percentage
allowed us to measure the squared visibility of Arcturus at spatial frequencies near 20 cycles/arcsec
as $0.49 \pm 0.26$. In combination with speckle data taken with the DSSI speckle camera at the ARC 3.5-m Telescope, we filled in the 
lower spatial frequencies for the star's visibility profile. Finally, by fitting a limb-darkened profile to the visibility
curve in the Fourier domain, we obtain a diameter for Arcturus of $21.5 \pm 1.8$ mas. While the uncertainty is larger than
the Michelson-style long baseline optical interferometers have obtained, this is the first measurement of 
Arcturus' diameter that incorporates some intensity interferometry data. The speckle dataset, supplemented by the SCSI
result, yields a 
measurement that is similar in precision to other stellar diameters obtained with modern intensity interferometers to date.
(We measure the diameter to 8\% uncertainty.)
Our measurement is also consistent with those of previously published studies on Arcturus. 

\begin{acknowledgements}
We gratefully acknowledge support from the National Science Foundation in the completion of this work, specifically
grants AST-1429015 and AST-1909582. We are also grateful to the Connecticut Space Grant Consortium for supporting
our 2025 observing campaign with SCSI through Faculty Research Grant P-2332, and to the Board of Regents of the 
Connecticut State University system for a Connecticut State University Research Grant.  Finally, we thank the excellent staff at 
Apache Point Observatory for their help in securing the speckle observations, and the anonymous referee for helpful comments during 
the review process.

This work presents results from the European Space Agency (ESA) space mission Gaia. Gaia data are being processed by the Gaia Data Processing and Analysis Consortium (DPAC). Funding for the DPAC is provided by national institutions, in particular the institutions participating in the Gaia MultiLateral Agreement (MLA). The Gaia mission website is https://www.cosmos.esa.int/gaia. The Gaia archive website is https://archives.esac.esa.int/gaia. We also made use of the SIMBAD database, operated at CDS, Strasbourg, France. 

\noindent
{\it Facilities:} APO:ARC.
 \end{acknowledgements}


\bibliography{horch_aj31a}{}
\bibliographystyle{aasjournalv7}




\end{document}